\documentclass[fleqn,usenatbib]{mnras}
\usepackage[T1]{fontenc}
\usepackage{ae,aecompl}
\usepackage{graphicx}   
\usepackage{amsmath}    
\usepackage{amssymb}    
\usepackage{float}

\newcommand{\nntab}[2]{ \multicolumn{2}{#1}{#2} }
\newcommand{\Gaia}{{\it Gaia}}
\usepackage{xcolor}
\definecolor{Dred}{rgb}{0.312,0.070,0.070}
\definecolor{Dblue}{rgb}{0.070,0.070,0.312}

\newcounter{note}
\let\oldmarginpar\marginpar
\renewcommand\marginpar[1]{\-\oldmarginpar[\raggedleft\footnotesize #1]%
{\raggedright\footnotesize #1}}

\newcommand{\Frac}[2]{\frac{\displaystyle\strut #1}{\displaystyle\strut #2} }
\newcommand{\vex}{\vspace{1ex}}
\newcommand{\dss}{\displaystyle}
\newcommand{\tss}{\textstyle}

\newcommand{\der}[2] {\frac{ \partial #1 }{ \partial #2 }}

\newcommand{\dintinf}{\int\limits_{\enskip-\infty}^{\enskip+\infty}\hspace{-0.90em}\int}

\newcommand{\lp}{ \left(  }
\newcommand{\rp}{ \right) }

\renewcommand{\Re} { {\rm Re}\hspace{0.2em} }
\renewcommand{\Im} { {\rm Im}\hspace{0.2em} }
\newcommand{\Ojpi}{\mbox{$O_\mathrm{j}$+$i$}}
\newcommand{\Ojpd}{\mbox{$O_\mathrm{j}$+$d$}}
\newcommand{\Ojpz}{\mbox{$O_\mathrm{j}$+$0$}}
\newcommand{\Ojmi}{\mbox{$O_\mathrm{j}\!-\!i$}}
\newcommand{\Ojmd}{\mbox{$O_\mathrm{j}\!-\!d$}}
\newcommand{\Ojmz}{\mbox{$O_\mathrm{j}\!-\!0$}}

\newcommand{\timezone}{-0400}
\newcommand{\Number}[1]{\ifnum#1<10\relax0\number#1\else\number#1\fi}
\newcommand{\isodate}{
\count151=\time
\divide\count151 by 60
\count151=\count151
\multiply\count151 by 60
\count152=\time
\advance\count152 by -\count151
\divide\count151 by 60
\count152=\count151
\multiply\count151 by 60
\count153=\time
\advance\count153 by -\count151
\Number{\year}.\Number{\month}.\Number{\day}--\Number{\count152}:\Number{\count153} \enskip \timezone
}
\definecolor{OliveGreen}{rgb}{0,0.6,0}

\title[Consequencies of AGN optical mas-scale structure]
      {\vspace{-4ex} \LARGE\bf Observational consequences of optical band milliarcsecond-scale 
       structure in active galactic nuclei discovered by {\it Gaia}}
\author[Petrov and Kovalev]{
L.~Petrov$^{1,2}$\thanks{E-mail: Leonid.Petrov@lpetrov.net},
and
Y.~Y.~Kovalev$^{2,3,4}$\\
$^1$Astrogeo Center, 7312 Sportsman Dr., Falls Church, VA 22043, USA\\
$^2$Moscow Institute of Physics and Technology, Dolgoprudny, Institutsky per., 9, Moscow, Russia \\
$^3$Astro Space Center of Lebedev Physical Institute, Profsoyuznaya 84/32, 
   117997 Moscow, Russia\\
$^4$Max-Planck-Institut f\"ur Radioastronomie, Auf dem H\"ugel 69, 
   53121 Bonn, Germany
}

\date{Accepted 2017 July 10. Received 2017 June 26; in original form 2017 April 21}

\pubyear{2017}

\begin{document}
\volume{474}
\pubyear{2017}
\setcounter{page}{3775}
\label{firstpage}
\pagerange{3775--3787}

\maketitle

\begin{abstract}
   We interpret the recent discovery of a preferable VLBI/\Gaia\ offset 
   direction for radio-loud active galactic nuclei (AGNs) along the 
   parsec-scale radio jets as a manifestation of their optical structure on 
   scales of 1 to 100 milliarcseconds. The extended jet structure affects 
   the \Gaia\ position stronger than the VLBI position due to the difference 
   in observing techniques. \Gaia\ detects total power while VLBI measures 
   the correlated quantity, visibility, and therefore, sensitive to 
   compact structures. The synergy of VLBI that is sensitive to the position 
   of the most compact source component, usually associated with the opaque radio 
   core, and \Gaia\ that is sensitive to the centroid of optical emission, opens
   a window of opportunity to study optical jets at milliarcsecond resolution, 
   two orders of magnitude finer than the resolution of most existing optical 
   instruments. We demonstrate that strong variability of optical jets is 
   able to cause a jitter comparable to the VLBI/\Gaia\ offsets at a quiet state, 
   i.e. several milliarcseconds. We show that the VLBI/\Gaia\ position jitter 
   correlation with the AGN optical light curve may help to locate the 
   region where the flare occurred, estimate its distance from the super-massive black 
   hole and the ratio of the flux density in the flaring region to the total 
   flux density.
\end{abstract}


\section{Introduction}

  The European Space Agency \Gaia\ project made a quantum leap in the 
area of optical astrometry. The secondary dataset of the first data release 
(DR1) contains positions of 1.14 billion objects \citep{r:gaia_dr1}
with median uncertainty 2.3~mas. Although the vast majority of \Gaia\ detected
sources are stars, over one hundred thousands of extragalactic objects, mainly
active galactic nuclei (AGN), were also included in the catalogue. The only 
technique that can determine positions of AGNs with comparable accuracy 
is very long baseline interferometry (VLBI). The first insight on comparison 
of \Gaia\ and VLBI position catalogues \citep{r:gaia_icrf2,r:gaia1} 
revealed that the differences in VLBI/\Gaia\ positions are close to reported 
uncertainties, though a small fraction of sources ($\sim\! 6$\%) show significant 
offsets. We will call these sources genuine radio optical offset (GROO) objects.

\begin{figure}
   \centering
   \includegraphics[width=0.47\textwidth]{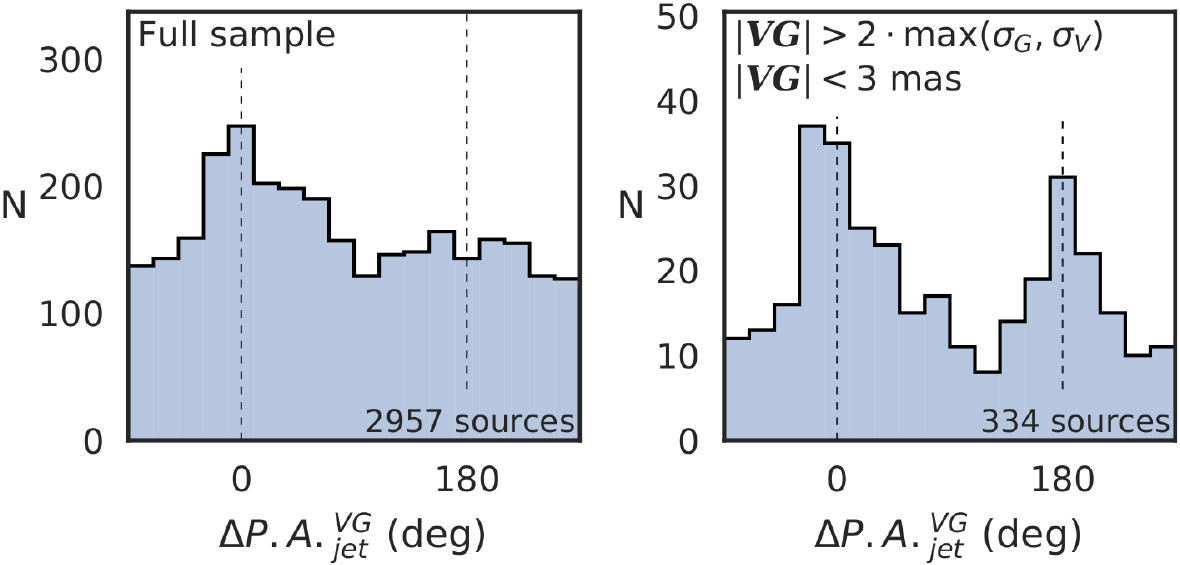}
   \caption{
            Histograms of direction vectors of VLBI/\Gaia\ offsets with
            respect to the jet directions. The vertical dashed lines correspond
            to a case when the direction of the Gaia position offset with
            respect to the VLBI position is along the jet direction ($0^\circ$) 
            and opposite to the jet direction ($180^\circ$). The left plot shows 
            the distribution for the full sample of 2957 VLBI/\Gaia\ matches 
            with the probability of false association less than $2 \cdot 10^{-4}$ 
            and with reliably determined jet directions. The right plot shows 
            the histogram  for the sub-sample of 334 sources with offsets that 
            a)~are shorter than 3~mas, and b)~longer than the maximum of 
            both $2\sigma$ VLBI and \Gaia\ position uncertainties. 
            The~Figure is reproduced from~\citet{r:gaia2} with permission from 
            Astronomy \& Astrophysics, (c) ESO.
           }
   \label{f:hist}
\end{figure}

  We presented argumentation in \citet{r:gaia1} that unaccounted 
systematic errors or blunders in analysis of VLBI or \Gaia\ data can
explain offsets for some sources, but cannot explain offsets for the majority
of GROO objects. Further analysis of \citet{r:gaia2} revealed that 
VLBI/\Gaia\ offsets of a general population of radio-loud AGNs, not 
only the matching sources with statistically significant offsets, have 
a preferable direction along the jet that is detected at milliarcsecond scale 
for the majority of radio sources (see Fig.~\ref{f:hist}). The existence 
of the preferable direction that is highly significant completely rules out 
alternative explanations of VLBI/\Gaia\ offsets as {\it exclusively} due 
to unaccounted errors in VLBI or \Gaia\ positions. Such errors, if exist, 
should cause either a uniform distribution of radio/optical position offsets, 
or have other preferable directions, for instance, across the declination 
axis (atmosphere-driven systematic errors in VLBI) or along the 
predominant scanning direction (\Gaia\ systematic errors). The preferable 
direction along the jet (Fig.~\ref{f:hist}) can be caused only by the intrinsic 
core-jet morphology. Our Monte Carlo simulation \citep{r:gaia2} showed 
that either offsets in the direction along the jet should have the mean 
bias exceeding 1.2~mas or the distribution of offsets should have the 
dispersion exceeding 2.6~mas in order to explain the histogram in Fig.~\ref{f:hist}.
We should emphasize that two factors resulted in a detection of a
preferable direction of VLBI/\Gaia\ offsets: a large sample of matches and 
measurement of jet directions at milliarcsecond scales, which corresponds
to parsec distances. In general, jet directions at arcsecond scales 
(kiloparsec distances) are significantly different from directions at milliarcsecond
scale \citep[See Fig.~6 in][]{r:kharb10}. Analyzing a small sample 
of VLBI/\Gaia\ matches and jet directions at arcsecond scales does not 
permit to reveal the systematic pattern as it was demonstrated 
by \citet{r:mak17}.

  There are two known systematic effects that can cause a bias in VLBI 
positions along the jet direction and thus, contribute to the observed 
pattern of VLBI/\Gaia\ position offsets at $180^\circ$ of the jet direction.
The true jet origin, the region at the jet apex, is thought to be invisible 
to an observer. It is opaque and has optical depth $\tau\gg1$ due to synchrotron 
self-absorption. The jet becomes visible further away from the origin when 
optical depth reaches $\tau\approx1$ at the apparent jet base, we call this 
region the core. The higher the frequency, the closer the observed core 
to the jet apex \citep[e.g.,][]{Kovalev_cs_2008,Sullivan09_coreshift,pushkarev_etal12,Sokol_cs2011,kutkin_etal14,r:1030,r:3C273_lisakov}.
This effect is called the core-shift. \citet{Kovalev_cs_2008} 
predicted that the apparent jet base in optical band will be shifted at 0.1~mas level 
with respect to the jet base at 8~GHz opposite to the jet direction because 
of frequency dependence of the core-shift. However, when the core-shift 
depends on frequency as $f^{-1}$, it has zero contribution to the 
ionosphere-free linear combination of group delays that is used for absolute
VLBI astrometry \citep{Porcas_cs2009} and thus, does not affect the absolute
VLBI positions. The \citet{r:bla79} model of a purely synchrotron 
self-absorbed conical jet in equipartition predicts the core-shift 
dependence on frequency $f^{-1}$. Observations \citep[e.g.][]{Sokol_cs2011} 
show no systematic deviation from this frequency dependence. The residual 
core-shift for the objects with the core-shift frequency dependence 
different than $f^{-1}$ \citep[e.g.,][]{kutkin_etal14,r:3C273_lisakov} 
is over one order of magnitude too small to explain Fig.~\ref{f:hist}. 
In addition to the synchrotron self-absorption, an external 
absorption of the jet base can happen in the broad-line region or 
the dusty torus. It strongly depends on jet orientation 
\citep[e.g.,][]{UP95}. It might further shift VLBI and/or \Gaia\ 
positions along the parsec-scale jet in case if emission of 
the jet is significant.

  The second effect is the contribution of the asymmetric radio structure 
to group delay that is commonly ignored in VLBI data analysis due to 
complexity of its computation. As we will show later, the median bias
in source position caused by the neglected source structure contribution
is below 0.1~mas at 8~GHz, which is also too small to explain 
the histograms in Fig.~\ref{f:hist}.

  The remaining explanation of the observed preferential direction of 
VLBI/\Gaia\ offset at $0^\circ$ of the jet direction is the presence of optical 
structure of AGNs on scales below the \Gaia\ point-spread function (PSF) that, 
according to \citet{r:gaia_fabricious16}, has the typical full width half 
maximum (FWHM) around $100 \times 300$~mas. Since at the moment there 
does not exist an instrument that could produce direct optical images 
at milliarcsecond resolution of objects of 15--20 magnitude, the proposed 
explanation can be supported only by indirect evidence. 

  This motivated us to consider the problem in detail and answer four 
questions. 1)~Can the small population of known optical AGN jets at separations 
$0.2''$--$20''$ be considered as a tail of the broader population of optical 
jets? 2)~What are the consequences of the presence of optical AGN jet structure
at scales 1--200~mas that can be verified or refuted by future observations? 
3)~What kind of insight to AGNs physics can provide us these observational 
consequences? 4)~How does the presence of optical structure affects the 
stability of AGN \Gaia\ positions and how to mitigate them? The layout of 
the subsequent discussion follows this logic.

  We use the following naming convention. The ``core'' is the apparent base 
of an AGN jet; its position is frequency dependent due to synchrotron 
self-absorption of the true base and is expected to appear further down 
the AGN jet with increasing observing wavelength; and the ``jet'' is the 
rest of the AGN jet structure.

\section{Impact of optical jets on source position}

   As the term ``active galactic nucleus'' suggests, super-massive black holes 
(SMBHs) are assumed to be at rest in the nuclei of their host galaxies because 
dynamical friction against the surrounding stars and gas will eventually make 
an offset SMBH in an isolated galaxy sink to the bottom of the host galaxy 
gravitational potential. In the absence of strong interaction with companion 
galaxies, the SMBH position will coincide with the center of mass of the star 
population of the host galaxy. \Gaia\ measures positions of the source's 
centroid. In the absence of asymmetric structures, such as optical jets, 
the position of the centroid in general coincides with the position of the 
SMBH and therefore, the \Gaia\ position will match to the VLBI position of the 
core that is located in the vicinity of the SMBH. Recent galaxy mergers with SMBHs 
may produce massive stellar bulges containing two or more SMBHs temporarily 
offset in position and velocity. Extensive searches of such binary AGNs that 
exhibit parsec-scale radio emission revealed only two objects 
\citep{r:0402+379,r:bingo1} that have been firmly confirmed with VLBI 
observations. Thus, such objects are rare.

  If the optical jet or its part is confined within the \Gaia\ PSF, 
its contribution changes the position of the centroid $C_x$
along direction $x$:
\begin{eqnarray}
   \begin{array}{lcl}
       C_x  & = & \Frac{\int I(x) \, w(x-x_0) \, x \, dx}
                       {\int I(x) \, w(x-x_0) \, dx},       
   \end{array}
   \label{e:e1}
\end{eqnarray}
  where $I(x)$ is the intensity distribution along axis $x$ and $w(x-x_0)$ is 
a weighting function normalized to unity --- a projection of the PSF 
to the direction $x$. Since the centroid depends linearly on spatial 
coordinates, the presence of the jet shifts the position of the centroid 
with respect to the core at 
\begin{eqnarray}
   \begin{array}{lcl}
       C_x  & = & \Frac{\int I^\mathrm{j}(x) \, w(x-x_0) \, x \, dx} 
                       {\int I^\mathrm{j}(x) \, w(x-x_0) \, dx \: + \: 
                       \int I^\mathrm{r}(x) \, w(x-x_0) \, dx} ,
   \end{array}
   \label{e:e2}
\end{eqnarray}
  where $I^j(x)$ is the jet intensity distribution and $I^r(x)$ is the remaining 
intensity distribution after jet subtraction. If the jet can be presented 
as a sum of delta-functions, and neglecting $w(x_k - x_0) - 1$, 
which corresponds to a case when $x_k$ is significantly less than PSF FWHM, 
the expression \ref{e:e2} is reduced to
\begin{eqnarray}
   \begin{array}{lcl}
       C_x  & = & \dss\sum\limits_{k} x_k \: \Frac{F^\mathrm{j}_k}{F^\mathrm{j}_k + F^\mathrm{r}_k},
   \end{array}
   \label{e:e3}
\end{eqnarray}
   where $F_k$ is the flux density of the $k$-th delta-function at the 
position $x_k$ and $F^\mathrm{r}_k$ is the remaining flux density excluding 
the $k$-th delta-function. 

\begin{figure}
\centering
  \includegraphics[width=0.48\textwidth]{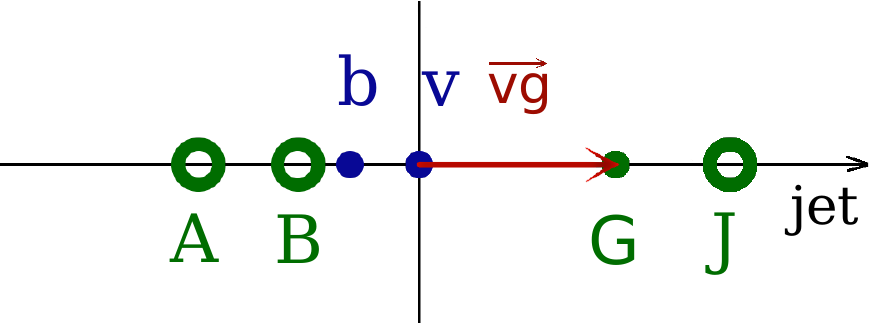}
  \caption{
           A simplified diagram of the AGN structure. The VLBI position is 
           shown with '{\sf v}'. It is shifted with respect to the apparent 
           VLBI jet base $b$ (the radio core) at a given frequency due to 
           unaccounted radio source structure contribution to its position
           estimate in the direction along the jet. The optical centroid 
           '{\sf G}' is a superposition of the emission from the accretion
           disk '{\sf A}', apparent \Gaia\ jet base (the optical core) '{\sf B}',
           and optical jet '{\sf J}'. The accretion disk is expected to be 
           very close to the optical core. The optical jet may be absent. 
           Astrometric observations provide us the VLBI/\Gaia\ offset 
           $\vec{\mbox{vg}}$ while VLBI imaging allows us to measure the 
           radio parsec-scale jet direction. 
           }
   \label{f:dia}
\end{figure}

  Fig.~\ref{f:dia} shows schematically an AGN milliarcsecond-scale structure. 
The accretion disk associated with an SMBH '{\sf A}' does not necessarily 
coincides with the core and may be shifted with respect to the jet base. 
However, radio images that show the counter-jet set the limit on its displacement 
with respect to the jet base to a fraction of a milliarcsecond. We assume 
that the SMBH is located at the center of mass of a galaxy and the centroid of 
the hosting galaxy starlight coincides with the center of mass. This condition may 
not be always fulfilled in the presence of dust. The contribution of the 
coreshift to the VLBI position derived from dual-band radio observations, 
the frequency-dependent vector $\vec{bv}$, is limited to the deviation of the 
coreshift dependence on frequency from $f^{-1}$. According to results of 
\citet{Sokol_cs2011}, it is mostly below 0.1~mas. The contribution of source 
structure, being ignored, may cause a bias in the estimate of the position of 
the apparent jet base '{\sf b}' along the jet direction. Point '{\sf J}' in 
the diagram shows the centroid of an optical jet. 

  We do not have direct evidence that the jet base is displaced with 
respect to the accretion disk, but the estimates of the upper limits of 
such displacements mentioned above show that this is not the dominant 
contributor to the observed displacements. In accordance with this scheme,
in general, the centroid of optical emission is determined by 
four parameters: flux density of the starlight $F_\mathrm{s}$ computed 
by integration of its intensity distribution; flux density of the optical 
core $F_\mathrm{c}$; flux density of the optical jet $F_\mathrm{j}$ 
produced by integration of its intensity distribution $I_\mathrm{j}$ and 
the displacement of its centroid with respect to the SMBH $d_\mathrm{j}$
({\bf BJ} vector on the diagram). Note that in a case of large offsets 
of optical emission centroids '{\sf G}', say greater than 1~mas, we can 
neglect the hypothetical displacement of the optical core '{\sf B}' with 
respect to the SMBH location '{\sf A}', $d_\mathrm{c}$. In that case, 
the displacement of the optical image centroid with respect to the SMBH 
is determined by two parameters: 
$r_\mathrm{j}=F_\mathrm{j}/(F_\mathrm{j} + F_\mathrm{s} + F_\mathrm{c})$ 
and $d_\mathrm{j}$. According to expression \ref{e:e3}, 
$C_x = r_\mathrm{j} \, d_\mathrm{j}$. As we will show below, 
applying data reduction that exploits radio source images, we can 
determine position of point '{\sf B}' with VLBI. Then, ignoring 
the shift of the starlight centroid and the optical core with respect 
to the SMBH, the difference VLBI/\Gaia\ will be equal to $C_x$.

\section{Known large optical jets}

  There are about two dozens of sources for which optical jets are detected
in images with separations of 1--$20''$ from galactic nuclei 
\citep[f.e.,][]{r:meyer17}. Since the jets are relatively weak, we can 
see them mainly in the sources that are at closer distances than the rest of 
the population. Besides, for the sources that are farther away, 
the angular separation of a jet from a nucleus will be smaller for a given 
linear separation. Jets at separations 1--$20''$ from nuclei are not expected 
to affect \Gaia\ positions since such separations are greater than the PSF. 
At the same time, it is instructive to get a rough estimate of how far the 
centroid would be shifted if sources with known optical jets were located 
at distances at which the jets would have been confined within the \Gaia\ PSF. 
We considered three sources, 3C264, 3C273, and M87, for which we found jet 
photometry in the literature. 

\begin{figure}
   \centering
   \includegraphics[width=0.475\textwidth]{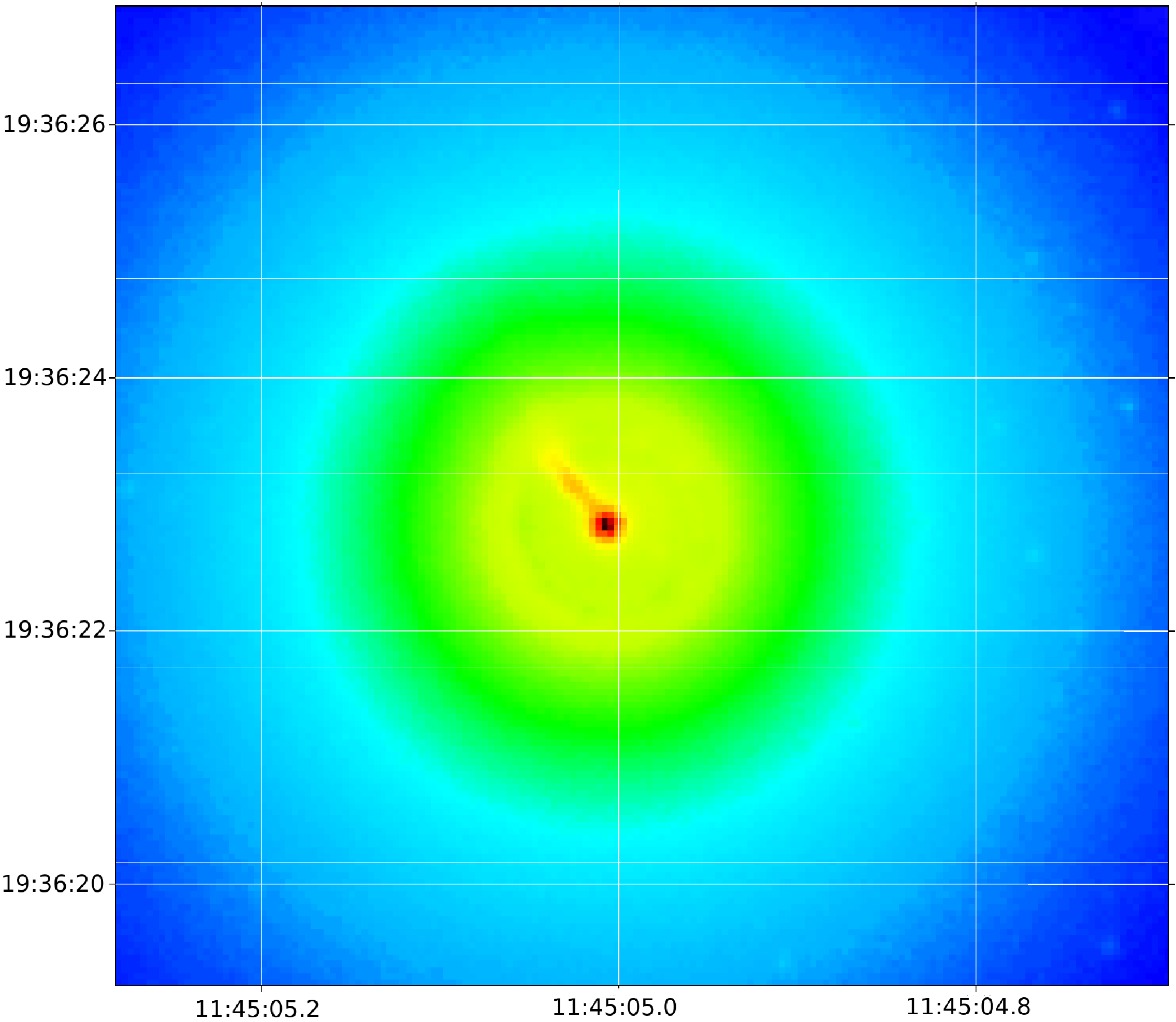}
   \caption{
            The archival {\it HST} image of 3C264 at 606~nm, 
            {\it HST} project ID 13327 \citep{r:3c264_image}.
           }
   \label{f:3c264}
\end{figure}

  3C264 (NGC\,3862, J1145+1936) is located at $z=0.0216$ and has a known
optical jet that is extending up to $0.8''$. Using photometry of the optical 
jet of 3C264 presented by \citet{r:3c264}, we got the estimates of the 
contribution of visible jet to the centroid: 15.6~mas. Independently, we used 
the archival Hubble Space Telescope ({\it HST}) image with the ACS/WFC 
instrument at 606~nm observed on August 21, 2015 (see Fig.~\ref{f:3c264}) 
and computed the differences in the centroid position within the area $0.15''$ 
around the core and within the whole image. The centroid difference was 
14.7~mas. At $z=0.067$ this optical jet would not have been resolved by 
the {\it HST}, but being confined in the \Gaia\ PSF, it would have caused 
a centroid shift of 5~mas.

  3C273 (J1229+0203) is located at $z=0.158$ and has the optical jet that 
is traced to $22''$. Using the photometry of \citet{r:3c273b}, we found that 
the contribution of the visible part of the jet to the centroid is 19~mas. 

  M87 (J1230+1223) at $z = 0.0046$ has a rich jet structure that is traced from 
distance of $0.8''$ up to $26''$. Using photometry of \citet{r:m87a} and 
\citet{r:m87b}, we found that the contribution of the visible part of the 
jet to the centroid is 56~mas. At $z=0.3$ the brightest components A, B, and 
C would be within $0.3''$ of the core and the contribution of the optical jet 
to the centroid position would be 1.2~mas.

  Examples of 3C264 and M87 show that if these sources be farther, at 
a distance that direct optical observations would not have been able to resolve
their jets, the shift of the centroid with respect to the core 
due to the presence of the jet would be several mas --- close to what 
VLBI/\Gaia\ comparison shows \citep{r:gaia2}. This does not prove our 
interpretation of the observed preference of the VLBI/\Gaia\ offset directions, 
but it demonstrates that properties of known optical jets permit such an 
interpretation. We hypothesize that the known extended jets are just the tail 
of the distribution with the bulk of optical jets being too short and too 
faint to be resolved from cores even at HST images. 

  In these examples we counted only a visible part of the jet at distances 
farther than 0.15~mas. A jet or its part with the centroid at 100~mas with 
respect to the SMBH and with the flux density at a level of 1\% of the total 
flux density shifts the \Gaia\ image centroid by 1~mas. \citet{r:perlman2010} 
present convincing argumentation that optical and radio emission is caused by 
the same synchrotron mechanism. Synchrotron emission in the radio range is 
traced from scales of ten microarcseconds to scales of arcminutes. Therefore, 
we conclude that the optical emission is not limited to scales of arcseconds 
where it could be detected with direct imaging but should be present 
at milliarcsecond scales as well.

\section{Impact of radio jets on source position}

  Comparison of optical jets with radio jets at arcsecond resolutions shows
that, in general, they are cospatial \citep[e.g.,][]{r:gabuzda2006}.
See also \citet{r:kharb10} for discussion of the misalignment between 
the pc-scale and kpc-scale jets in radio. The questions arises why the 
presence of the core does not shift VLBI and \Gaia\ positions the same 
way? There are three possible reasons.
First, starlight contributes in the optical range, but does not contribute
significantly in the radio range. For instance, if we subtract starlight, 
the contribution of the optical jet and the core would have shifted 
the centroid of M87 by 7--$9''$ \citep[computed using 
Table~1 in][]{r:m87b}. There is no evidence that the starlight 
can cause a shift of the optical centroid downstream the jet.
Second, since radio spectrum of a jet and a core are different, the ratio 
of the flux density that comes from the radio jet to the flux density that 
comes from the radio core extrapolated to the optical band should be 
different than in the radio range. Models of synchrotron 
parsec-scale jet emission \citep[e.g.,][]{r:mimica09} predict that regions 
downstream the apparent jet base have steep spectra. Assuming the same Doppler 
boosting, optical synchrotron jet emission is expected to have lower 
surface brightness than the radio one. 
Third, VLBI does not provide the position of the centroid. This requires 
further clarification.

 The response of a radio interferometer, the complex visibility function 
$V_{12}$, according to the Van~Zitter--Zernike theorem 
\citep{r:tms}, is
\begin{eqnarray}
   V_{12}(b_x,b_y,\omega) = e^{\tss i\omega\tau_0} 
         \hspace{-0.75em} \dintinf I(x,y,\omega)
         e^{\tss -i \, \omega (x b_x + y b_y )} dx\,dy
   \label{e:e4}
\end{eqnarray}
  where $\omega$ is the angular reference frequency of the received signal,
$\tau_0$ --- the geometric delay to the reference point on the source, and
$I$ --- the intensity distribution which depends on local Cartesian spatial 
coordinates with respect to the reference point in the image plane $x$, $y$, 
and frequency. $b_x$ and $b_y$ are the projections of the baseline vector 
$\vec{b} = \vec{r}_1 - \vec{r}_2$ between two stations $\vec{r}_1, \vec{r}_2$ 
to the plane that is tangential to the center of the map ($x=0$, $y=0$). 

  The observable used for determining source position is a group delay 
defined as 
\begin{eqnarray}
   \tau_{gr} = \der{}{\omega} \arg{V_{12}}.
   \label{e:e5}
\end{eqnarray}
  Typically, 10--100 estimates of group delay at different baselines 
at one or more epochs are used for deriving the source
position. Unlike a quadratic detector installed in the focal plane of 
an optical telescope, e.g., a CCD camera, each given estimate of group 
delay of an interferometer depends on the entire image in a substantially 
non-linear way. A response of an interferometer, the visibility function, 
is proportional to a harmonic of the spatial Fourier-spectrum of the image. 
VLBI observations provide the spatial spectrum sampled only in a limited 
range of harmonics. For typical observations used for deriving the source 
positions, the range of baseline vector projections to the source's tangential 
plane is 80--8000~km. This range of baseline vector projections according 
to the Fourier integral \ref{e:e4} corresponds to the range of 1--100~mas 
at the image plane when observations are made at 8~GHz. The interferometer 
is blind to spatial frequencies beyond that range due to a limited sampling 
of the visibility function. Features at the image smaller than that scale 
appear as point-like components. Features at the image larger than that scale, 
i.e. low surface brightness emission with variations beyond that scale, 
do not affect the visibilities at all.

The partial derivatives of group delay to source coordinates
\begin{eqnarray}
   \der{\tau_{gr}}{\alpha} = \frac{1}{c} \vec{b} \cdot \der{\vec{s}}{\alpha} + O(c^2)\,, \qquad
   \der{\tau_{gr}}{\delta} = \frac{1}{c} \vec{b} \cdot \der{\vec{s}}{\delta} + O(c^2)
   \label{e:e6}
\end{eqnarray}
  are proportional to the baseline vector length. Here $\vec{s}$ is unit
the vector of source coordinates. Therefore, despite the interferometer
sees a range of spatial frequencies, the sensitivity of the interferometer
to source coordinates is dominated by the longest baselines. At longest
baselines, the interferometer is sensitive to the finest features of an image
that is comparable to the resolution of an array. Extended features, even if they
are detected by an interferometer and show up at an image, provide very
small contribution to a source position estimate. Therefore, a position 
of an extended object derived from the analysis of interferometric 
observations is related not to a centroid defined by expression~\ref{e:e1},
but to a different point.

  The expression~\ref{e:e5} can be reduced to
\begin{eqnarray}
   \tau_{gr} = \tau_o + \tau_\mathrm{s},
   \label{e:e7}
\end{eqnarray}
  where, if we ignore dependence of source structure on frequency within 
the recorded band, the contribution of source structure to group delay 
$\tau_\mathrm{s}$ is expressed as
\begin{eqnarray}
   \begin{array}{r@{\hspace{-4em}}l}
      \tau_{s}(b_x,b_y) = \Frac{2\pi}{c|\tilde{V}|^2} \\ & \Biggl[ 
              \Re \tilde{V}(b_x,b_y) \; \Im \lp \nabla \tilde{V}(b_x,b_y) \rp^\top  \cdot 
              \lp b_x, b_y \rp \: - \\ & \phantom{\Biggl[ \Biggr.}
              \Im \tilde{V}(b_x,b_y) \;  
              \Re \lp \nabla \tilde{V}(b_x,b_y) \rp^\top \cdot \lp b_x, b_y \rp \Biggl]\,.
  \end{array}
   \label{e:e8}
\end{eqnarray}
  Here we denote the visibility without the geometric term as $\tilde{V}$,
i.e. $\tilde{V} = V_{12}(\tau_0=0)$.

  The term $\tau_\mathrm{s}$ has a complicated dependence on the source image 
that can be expressed analytically only for some simplest cases 
\citep{r:cha90}. There are two approaches for the treatment of the $\tau_\mathrm{s}$ 
term in data analysis. The first approach is to compute $\tau_\mathrm{s}$ 
using an image. In that case the position will be related to the reference 
point on the image that is explicitly chosen. The second approach is to set 
$\tau_\mathrm{s}=0$ during data reduction, which is equivalent to choosing 
$I(x,y) = \delta(x,y)$. Term $\tau_s$ in general is not proportional to
the partial derivatives of group delay with respect to source coordinates. 
Therefore, its omission is not equivalent to a shift in source positions and 
it will not be absorbed entirely by causing a bias in the source position 
estimates. Large residuals will be removed during the outlier elimination 
procedure; smaller residuals will propagate to the solution and affect 
source positions. This approach is up to now commonly adopted in all VLBI 
data analyses, including those used for deriving source position catalogues, 
since the contribution of the source structure usually does not dominate 
the error budget.

  The magnitude of the position bias caused by ignoring $\tau_\mathrm{s}$ depends 
on many factors, including the observation schedule that affects a selection of 
of the Fourier transform harmonics of the source brightness distribution 
contributing to $\tau_\mathrm{s}$. For demonstrating the magnitude of the source 
structure contribution, we reprocessed observing session BL229AA from the VLBA 
MOJAVE program \citep{r:lister16} observed on September 26, 2016. This 24-hour 
experiment was designed to get high fidelity images of 30~objects at 15.3~GHz. 
Most target sources have rich structure, i.e. the sample was biased towards the 
sources with significant $\tau_\mathrm{s}$. We performed two full data analysis 
runs of the BL229AA observing session: the first with $\tau_\mathrm{s}$ computed 
according to the expression \ref{e:e8} utilizing the images generated during 
processing this experiment by the MOJAVE team and made publicly 
available\footnote{Available from 
\href{http://www.physics.purdue.edu/MOJAVE}{http://www.physics.purdue.edu/MOJAVE}}
and the second with $\tau_\mathrm{s}$ set to zero. The reference point on the 
image was set to the image peak intensity pixel for these tests. Our analysis 
included fringe fitting, elimination of all outliers exceeding 3 times
weighted root mean squares of residuals (1.2\% observations) and estimation of 
model parameters that included station positions, the Earth orientation 
parameters, clock function for all stations, except the reference one, 
represented with B-splines of the 1st degree, residual atmospheric path delay 
in zenith direction for all sites, also represented with B-spline of the 1st 
degree, and source coordinates. The weighted root mean squares of postfit 
residuals was 19.8~ps for the solution that uses $\tau_\mathrm{s}$ computed 
from the images and 21.1~ps for the solution that set $\tau_\mathrm{s}$ to 
zero. Source position uncertainties were at a range of 40--120~$\mu$as. 
Table~\ref{t:bl229aa} shows the result sorted in increasing the contribution 
of source structure to source position.

\begin{table}
   \caption{The contribution of source structure to source position 
            estimates from processing BL229AA 15~GHz VLBA observing session
            of the MOJAVE program \citep{r:lister16}. The third column 
            shows the magnitude of the offset from the lowest to the 
            highest values and the fourth column shows the position angle 
            of the offset with respect to jet direction. 
            $PA_\mathrm{j}=0$ corresponds to the offset towards the jet 
            direction of the source position estimate from the solution
            with $\tau_\mathrm{s}$ applied with respect to the estimate 
            from the solution with $\tau_\mathrm{s}$ set to zero. The fifth 
            column shows the position of the image centroid with respect to 
            the location of the image maximum.
           }
   \centering
   \begin{tabular}{llcrc}
      \hline
      J2000        & B1950      & $|\vec{bv}|$ offset & $PA_\mathrm{j}$ & Centroid  \\
      name         & name       & (mas)  & (deg)      & (mas)     \\
      \hline
      J0825$+$6157 & 0821$+$621 &   0.01 & $ -76$ &   0.17    \\
      J0510$+$1800 & 0507$+$179 &   0.01 & $ -98$ &   0.07    \\
      J0259$+$0747 & 0256$+$075 &   0.03 & $-174$ &   0.16    \\
      J0309$+$1029 & 0306$+$102 &   0.03 & $-162$ &   0.10    \\
      J2152$+$1734 & 2150$+$173 &   0.03 & $ 114$ &   0.45    \\
      J0505$+$0459 & 0502$+$049 &   0.04 & $-157$ &   0.25    \\
      J1031$+$7441 & 1027$+$749 &   0.04 & $ 179$ &   0.10    \\
      J1603$+$5730 & 1602$+$576 &   0.04 & $  91$ &   0.33    \\
      J1848$+$3244 & 1846$+$326 &   0.04 & $-131$ &   0.68    \\
      J0854$+$2006 & 0851$+$202 &   0.04 & $ -76$ &   0.07    \\
      J0017$+$8135 & 0014$+$813 &   0.05 & $ 127$ &   0.17    \\
      J1551$+$5806 & 1550$+$582 &   0.05 & $ 123$ &   0.13    \\
      J0131$+$5545 & 0128$+$554 &   0.06 & $ 163$ &   1.05    \\
      J1835$+$3241 & 1833$+$326 &   0.06 & $-102$ &   0.76    \\
      J2042$+$7508 & 2043$+$749 &   0.06 & $-160$ &   0.47    \\
      J2301$-$0158 & 2258$-$022 &   0.08 & $ 122$ &   0.12    \\
      J0642$+$6758 & 0636$+$680 &   0.08 & $ 132$ &   0.13    \\
      J1553$+$1256 & 1551$+$130 &   0.08 & $  -9$ &   1.98    \\
      J2202$+$4216 & 2200$+$420 &   0.09 & $ 170$ &   0.92    \\
      J0925$+$3127 & 0922$+$316 &   0.09 & $-179$ &   0.91    \\
      J0214$+$5144 & 0210$+$515 &   0.09 & $-155$ &   0.47    \\
      J2016$+$1632 & 2013$+$163 &   0.10 & $ 105$ &   0.18    \\
      J0839$+$1802 & 0836$+$182 &   0.11 & $ 178$ &   1.56    \\
      J1925$+$1227 & 1923$+$123 &   0.12 & $  20$ &   0.06    \\
      J1145$+$1936 & 1142$+$198 &   0.12 & $ 149$ &   0.56    \\
      J1756$+$1535 & 1754$+$155 &   0.14 & $ -13$ &   0.19    \\
      J1719$+$1745 & 1717$+$178 &   0.19 & $-155$ &   0.22    \\
      J1421$-$1118 & 1418$-$110 &   0.22 & $   1$ &   0.01    \\
      J1229$+$0203 & 1226$+$023 &   0.51 & $ -67$ &   2.58    \\
      J1153$+$4036 & 1151$+$408 &   2.40 & $-157$ &   1.06    \\
      \hline
   \end{tabular}
   \label{t:bl229aa}
\end{table}

\begin{figure}
   \centering
   \includegraphics[width=0.47\textwidth]{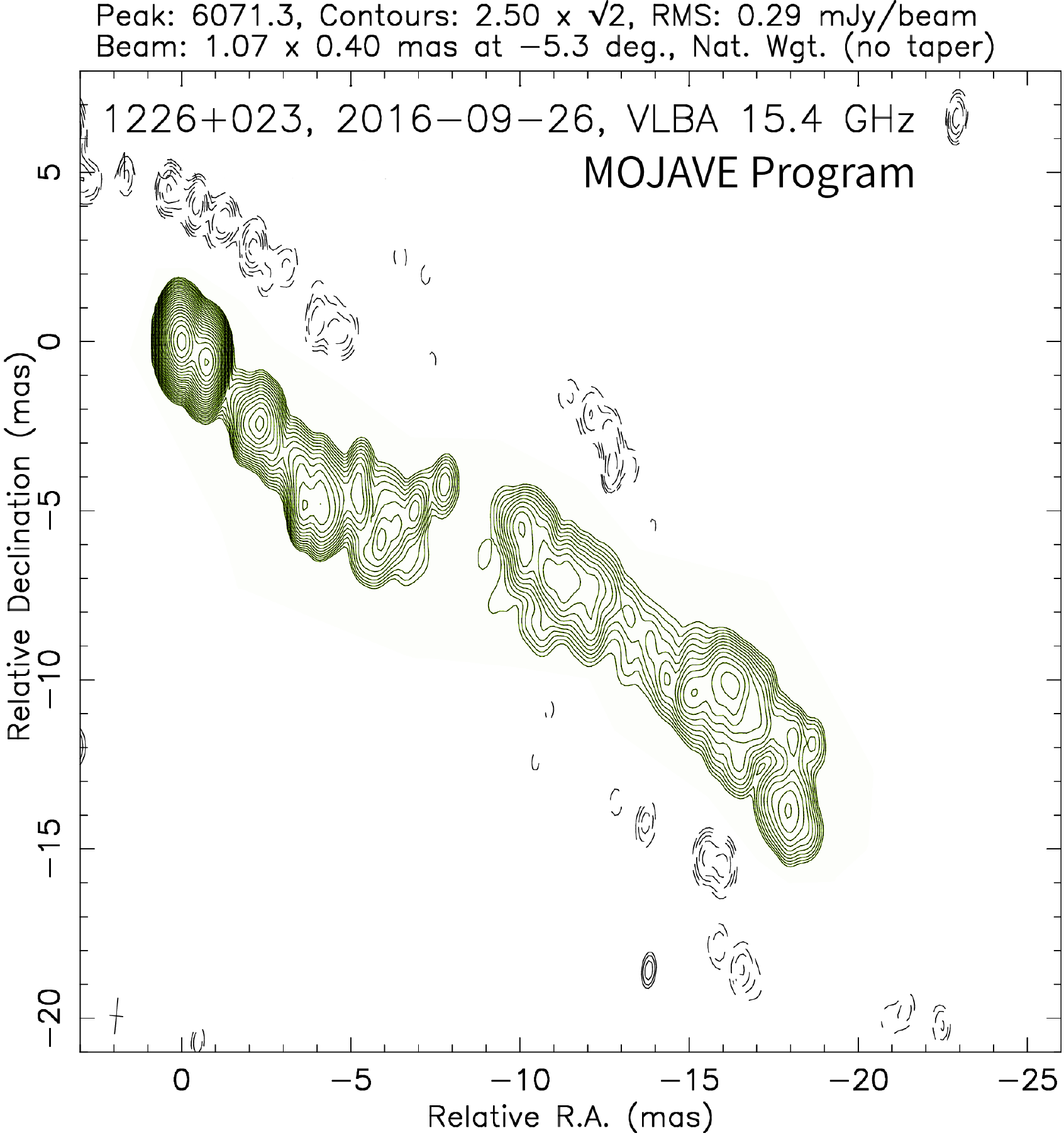}
   \par\vex\vex\par
   \includegraphics[width=0.47\textwidth]{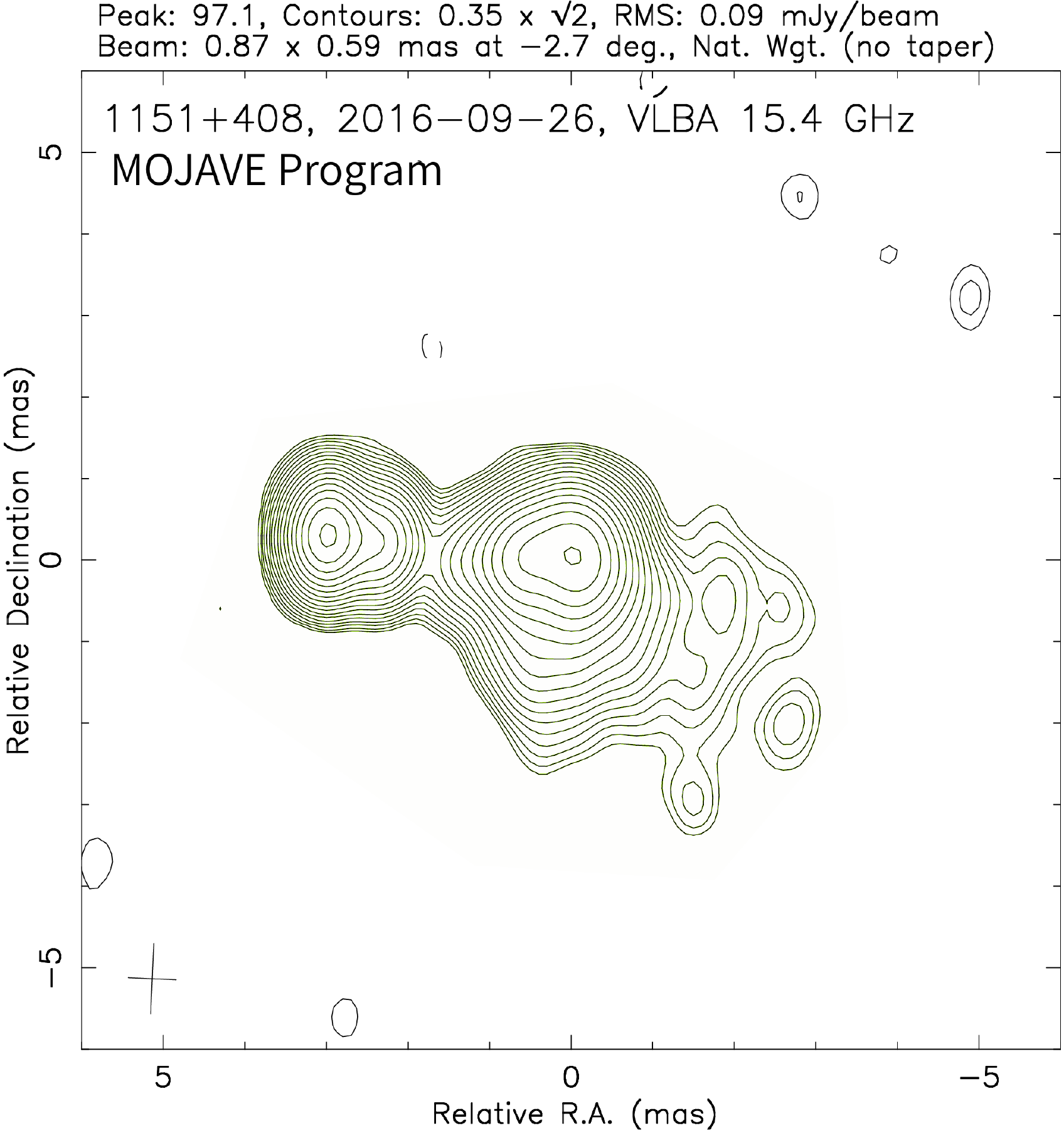}
   \caption{
            Images of the sources with the largest contribution of their
            structure to position estimates, 0.5~mas for J1229$+$0203 (3C273)
            and 2.4~mas for J1153$+$4036.
           }
   \label{f:bl229aa_images}
\end{figure}

  Analysis of the Table~\ref{t:bl229aa} shows that the median position bias
even for the sample of sources with rich structures is only 0.06~mas. It 
exceeds 0.5~mas only for two sources, J1229$+$0203 (3C273) and J1153$+$4036.
Their images are shown in Fig.~\ref{f:bl229aa_images}. In general, the 
sources with such structures are rare, less than 2\%. The position offset 
occurs predominately along the jet: either towards or opposite to the jet 
direction. The magnitude of the position offset has little in common with the 
magnitude of the shift of the centroid defined by expression~\ref{e:e1} 
with respect to the brightest component of the source.

  In order to illustrate further the effect of source structure on source
position from VLBI observations, we ran several simulations. We used conditions 
and the setup of VLBA observations of 3C273 within the BL229AA segment of the 
MOJAVE program and replaced the 3C273 image with a simulated image. Then we
repeated the procedure of outlier elimination and re-weighting and made two 
solutions with $\tau_\mathrm{s}$ computed from the simulated image and with 
$\tau_\mathrm{s}=0$ using exactly the same flagging and weights.

\begin{figure}
   \centering
   \includegraphics[width=0.232\textwidth]{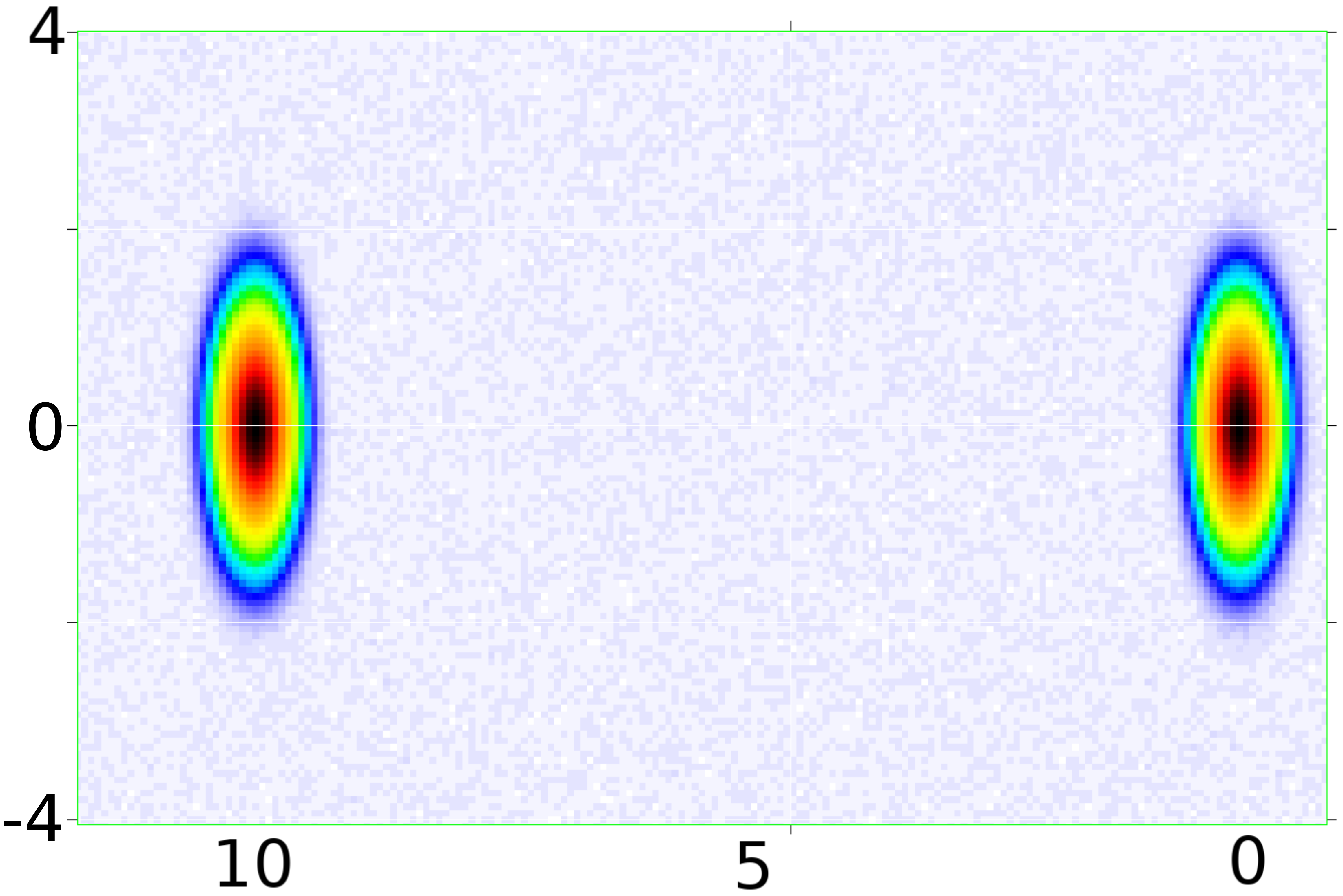}
   \hspace{0.00\textwidth}
   \includegraphics[width=0.232\textwidth]{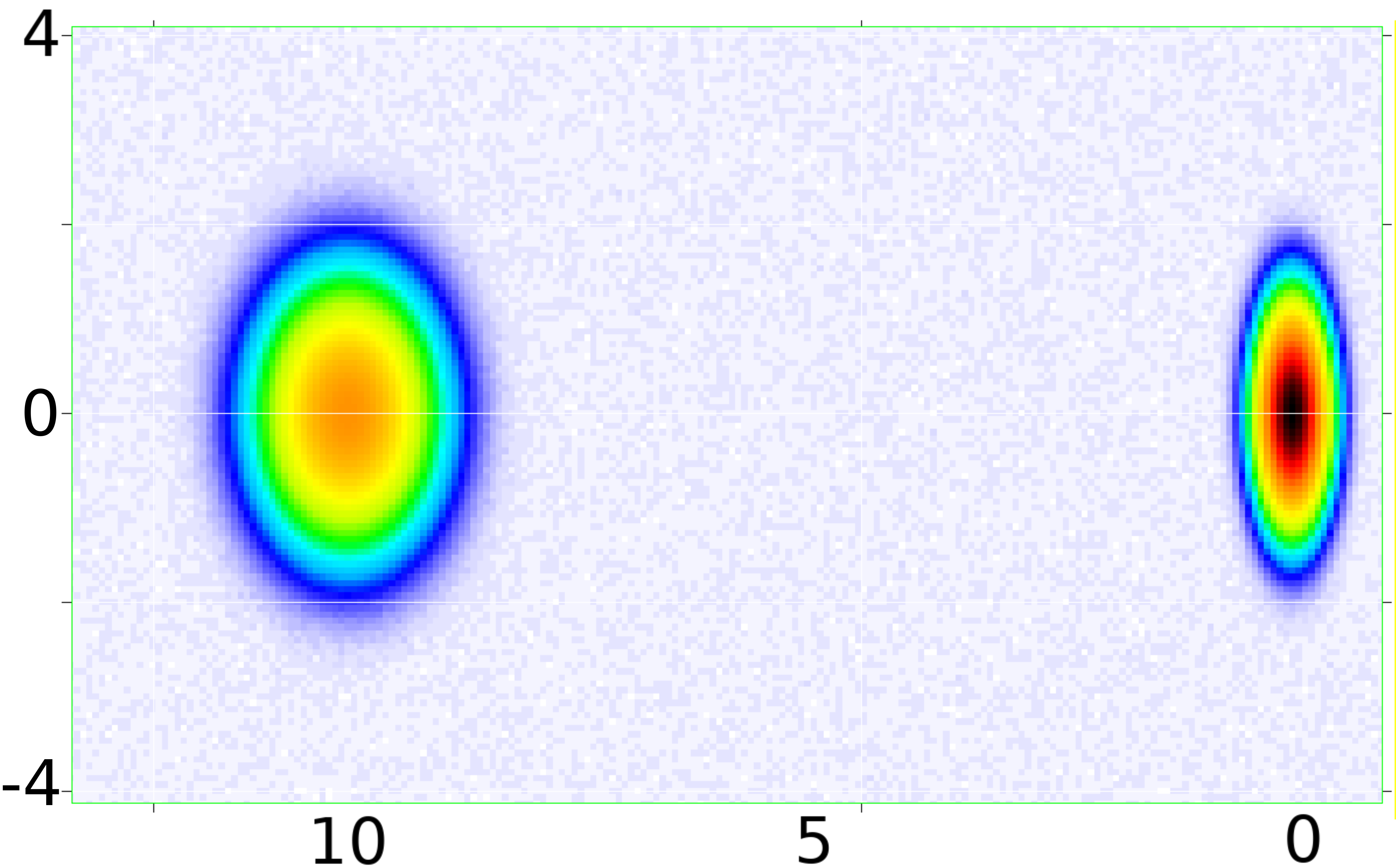}
   \includegraphics[width=0.232\textwidth]{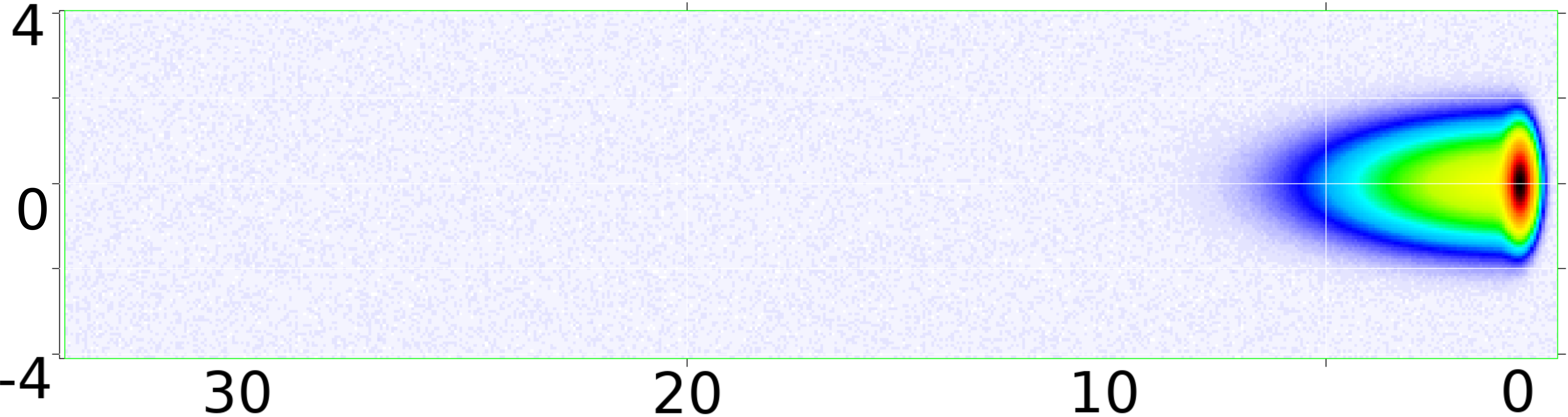}
   \hspace{0.00\textwidth}
   \includegraphics[width=0.232\textwidth]{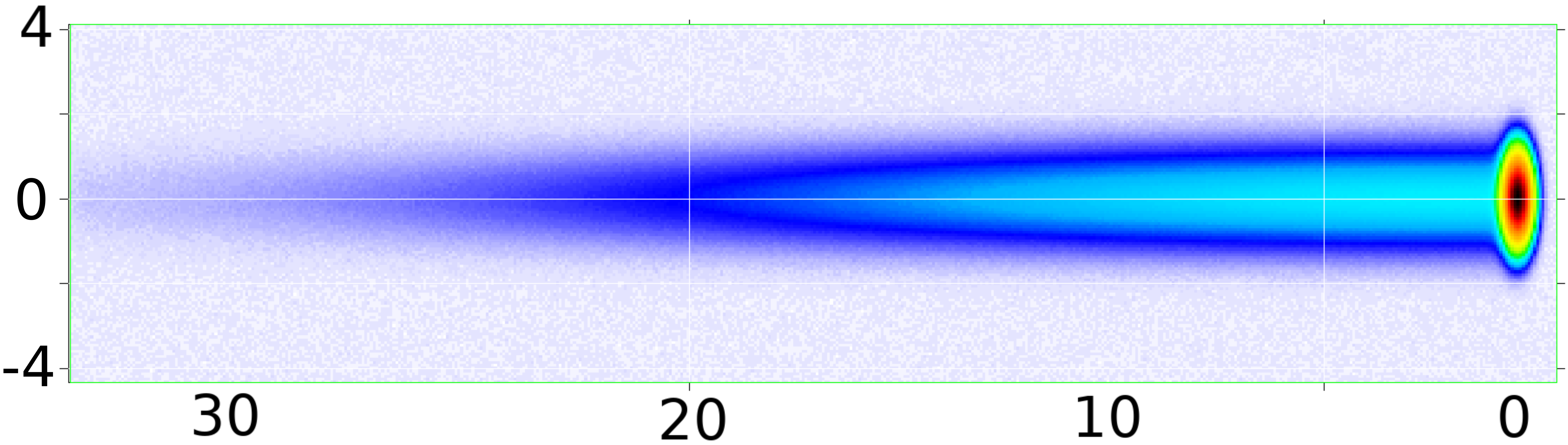}

   \par\vex\par
   \caption{
            Simulated maps for four cases. The maps are convoluted with
            the beam with FWHM axes $0.3 \times 1.0$ mas. Units along the axes are milliarcseconds.
           }
   \label{f:map_sim}
\end{figure}

  We modeled an image with two components, each with total flux density 1~Jy.
We considered four cases (See Fig.~\ref{f:map_sim}): 
\begin{enumerate}
   \item Both components are circular Gaussians with the FWHM 0.05 mas, 
         i.e. unresolved for BL229AA experiment. 
         The separation of components is 10 mas.

   \item The first component in the center of the field is a circular 
         Gaussian with the FWHM 0.05~mas, and the second displaced component 
         is a circular Gaussian with the FWHM 1.0~mas. 
         The separation of components is 10 mas.
         For comparison, the beam has FWHM size of $0.3 \times 1.0$~mas.

   \item The first component in the center of the field is a circular Gaussian
         with the FWHM 1.0 mas, and the second component is a one-sided 
         elliptical Gaussian at the same center as the first component and 
         the FWHM 1~mas along the declination axis and 5~mas along the right 
         ascension axis. The one-sided Gaussian is zero for $x<0.0$, i.e. 
         towards a decrease in right ascensions.

   \item The first component in the center of the field is a circular Gaussian
         with the FWHM 1.0 mas, and the second one is a one-sided elliptical 
         Gaussian at the same center with the FWHM 1~mas along the declination 
         axis and 30~mas along the right ascension axis.
\end{enumerate}


  Table~\ref{t:sim} shows estimates of the position offset of the solution
with $\tau_\mathrm{s}$ computed from the modeled image with respect to the solution
with $\tau_\mathrm{s}$ set to zero. The offset corresponds to the position bias caused 
by ignoring existing source structure. We see that 
only in a case when two components were equal unresolved Gaussians, the 
VLBI position estimate coincides with the centroid position. In all other 
cases the VLBI position estimate is very far from the centroid. The VLBI position 
estimate is sensitive to source structure mainly in a case when the second component 
has size less than the interferometer resolution. It may seem counter-intuitive 
that the presence of source structure perfectly aligned along the right 
ascension axis caused position offset along declination as well. In general,
$\tau_\mathrm{s}$ can only be partly recovered in estimates of source coordinates.
The remaining source structure contribution affects the parameter estimation 
process like noise. It propagates to the estimates of other parameters, including
declinations. We note that the contribution of actual jets to the position 
estimates would have been diluted even stronger since their typical shape is 
conical with the median apparent opening angle about 
$20^\circ$ \citep{r:pushkarev17}.

\begin{table}
   \caption{
            Results of simulation. The second and third columns show
            position estimate differences of the solution with $\tau_\mathrm{s}$ 
            computed from the simulated image with respect to the solution
            when $\tau_\mathrm{s}$ was set to zero. The fourth column shows the 
            displacement of the image centroid with respect to the 
            component right at the center of the simulated image.
           }
   \centering
   \begin{tabular}{lrrrr}
      \hline
       Case  & \nntab{c}{Offset estimates} & \nntab{c}{Centroid offsets}   \\
             & $\Delta \alpha$ & $\Delta \delta$ & $C_\alpha$ & $C_\delta$ \\
             &  mas           &  mas          &  mas     & mas             \\
      \hline
       1     &   5.000        &   0.0         &  5.000   & 0.000  \\
       2     &   0.302        &   0.100       &  5.000   & 0.000  \\
       3     &   0.153        &   0.003       &  0.857   & 0.000  \\
       4     &   0.260        &   0.068       &  4.989   & 0.000  \\
      \hline
   \end{tabular}
   \label{t:sim}
\end{table}

%

\section{Kinematics of AGN jets}

  Early VLBI observations revealed that source images are changing with 
time \citep{r:super_c}. Jet kinematics was extensively studied at 
both northern \citep[e.g.,][]{r:piner12,r:lister16,r:boston} and southern 
hemispheres \citep[e.g.,][]{r:tanami}. Here we provide a concise summary 
of the results relevant for our problem.

  The intensity of the jet emission changes with time. These changes are 
in general frequency dependent. The intensity distribution along the jet 
is not uniform. The apparent jet origin (the core) is usually the 
brightest feature. There are areas of stronger emission or weaker emission 
that may not be visible on an image due to its limited dynamic range. Jets 
are continuous and mostly have a conical shape. Their emission steadily 
decreases with the distance from the core. At the same time, some jet 
regions (or features, components, knots, blobs) might look brighter than 
the underlying jet. The components also dim and disappear with the 
distance to the core.
The jet direction is stable for over decades, although ejection angle 
of features may vary over several tens of degrees. The typical circular 
standard deviation in position angle of jet components is 
$\sim\! 10^\circ$ \citep{r:mojave_paper_x}. Jet components may appear 
at different parts of a jet, and typically show the radial 
motion \citep{r:lister16}.
Some jet components are observed to have non-radial motion 
\citep{r:lister16} but this does not affect the overall conical jet 
shapes especially for stacked multi-epoch multi-year images 
\citep{r:pushkarev17}. Moreover, the non-radial motion and bending 
accelerations tend to better align features with the inner 
jet \citep{r:mojave_paper_xii}. 

  According to \citet[Table 5]{r:lister16}, a typical angular speed 
of features in AGN jets at parsec scales found for the large MOJAVE sample 
is 0.1~mas\,y${}^{-1}$ or slower. Different components of the same jet 
move with approximately the same  characteristic speed that represents the true flow, 
suggesting that the observed speed of the jet is an intrinsic property of 
a source being related to the underlying flow speed \citep{r:mojave_paper_x}. 
It can rarely reach values higher than 1~mas\,y${}^{-1}$ for nearby objects. 
And the extreme example comes with the nearby jet in M87 which shows superluminal 
speed in both radio and optical band up to 25~mas\,y${}^{-1}$ 
\citep{r:biretta99,r:cheung07}. 

\begin{figure}
    \centering
    \includegraphics[width=0.475\textwidth]{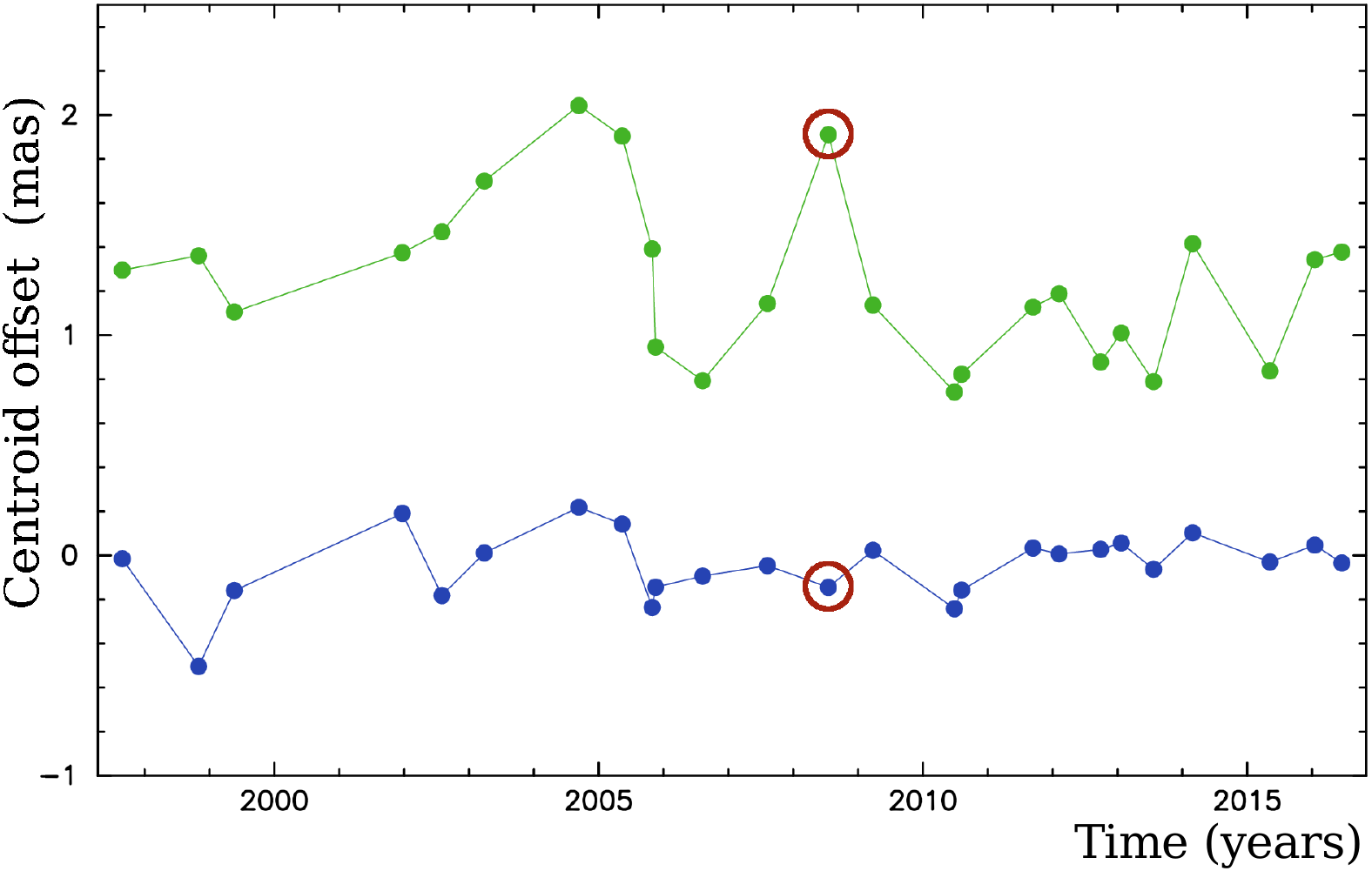}
    \par\vex\par
    \caption{
             Evolution of the centroid offset of J1829+4844 radio images 
             at 15.3 GHz with respect to the core. The green points 
             (upper part) show the centroid offsets along the jet 
             direction. The blue points (lower part) show the centroid 
             offsets transverse to the jet direction. The point 
             for the epoch of image in Fig.~\ref{f:J1829+4844_image} 
             is marked with a circle.
            }
    \label{f:J1829+4844_centroid}
\end{figure}

\begin{figure}
    \centering
    \includegraphics[width=0.475\textwidth]{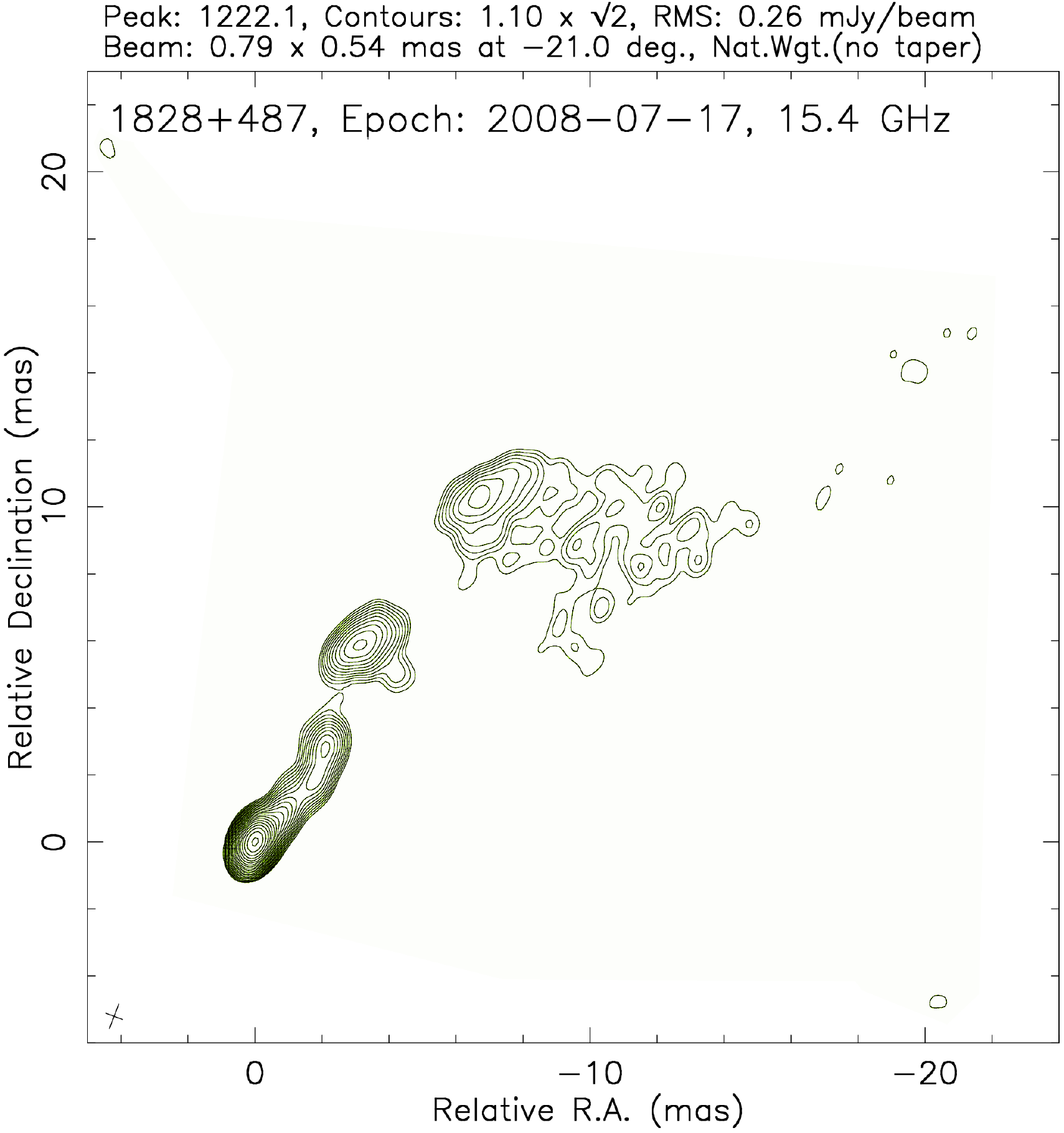}
    \par\vex\par
    \caption{
             Image of J1829+4844 --- the source with significant evolution 
             of its radio centroid (See Fig.~\ref{f:J1829+4844_centroid}).
            }
    \label{f:J1829+4844_image}
\end{figure}

\begin{figure*}
\centering
  \fbox{\includegraphics[width=0.22\textwidth]{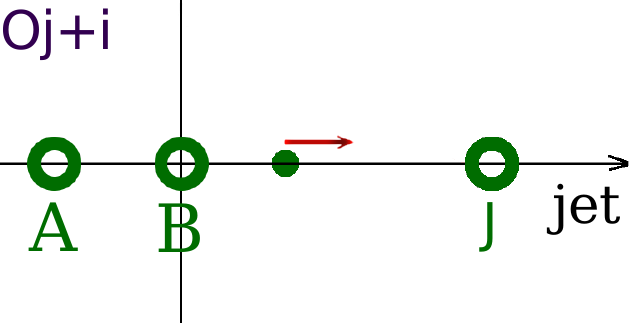}}\hspace{0.01\textwidth}
  \fbox{\includegraphics[width=0.22\textwidth]{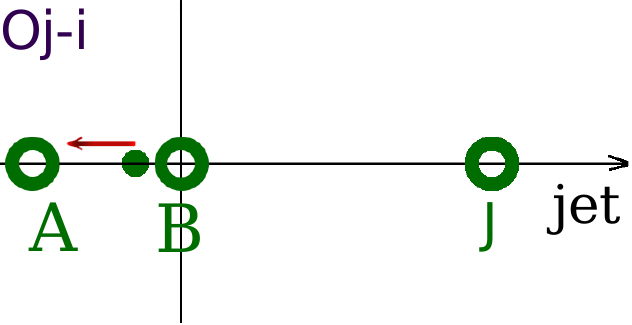}}\hspace{0.01\textwidth}
  \fbox{\includegraphics[width=0.22\textwidth]{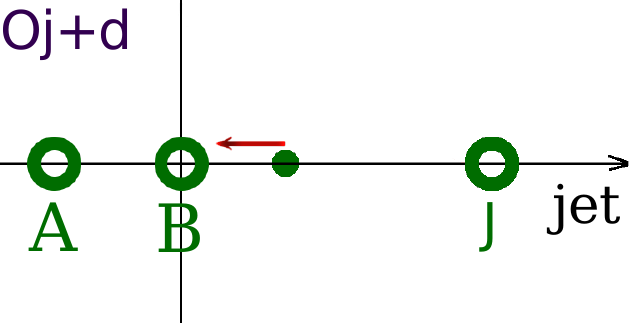}}\hspace{0.01\textwidth}
  \fbox{\includegraphics[width=0.22\textwidth]{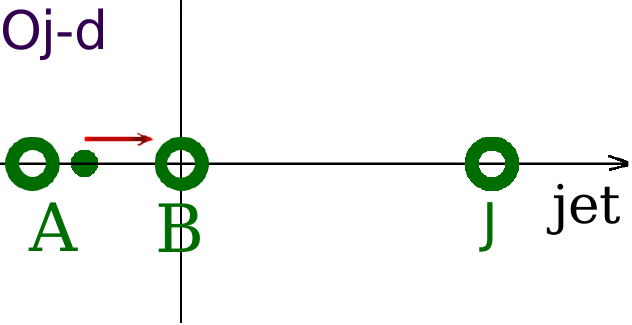}}
  \caption{A simplified diagram of the $O_\mathrm{j}$ projection changes 
           after a flare in the optical band: 
           1)~\Ojpi: positive projection decreases by modulo;
           2)~\Ojmi: negative projection increases by modulo;
           3)~\Ojpd: positive projection increases by modulo;
           4)~\Ojmd: negative projection decreases by modulo.
           The filled circle denotes the optical centroid.
           The labels are the same as in Fig.~\ref{f:dia}.
           }
   \label{f:dyn}
\end{figure*}

   Motion of bright components along the jet and changes of its flux 
density and the flux density of the core affect the position of the centroid. 
Fig.~\ref{f:J1829+4844_centroid} demonstrates changes of the centroid offset 
of radio image of J1829+4844 at 15.3~GHz (See its image in 
Fig.~\ref{f:J1829+4844_image}) with respect to the brightest feature 
that is associated with the radio core. We computed the centroid according to 
expression~\ref{e:e1} using images produced by the MOJAVE team from VLBA 
observations. We underline that the images, not the visibility data, were 
used in this analysis. The changes of the centroid offset due to the source 
structure evolution are over 1~mas peak-to-peak along the jet direction. 
As expected, images at epochs with low flux density level of the core emission 
tend to have higher offset and, opposite to that, a flaring core decreases the 
offset \citep[see the core modeling results in][]{r:mojave_paper_x}.
The root mean square (rms) of the centroid offset time series 
along the jet is 0.36 mas. The rms of the centroid offsets transverse the 
jet direction is 0.16~mas. We should note that, in general, centroid variations 
in optical and radio ranges are not expected to be the same since the relative 
weight of the core, the low surface brightness feature of jet, and the 
starlight are different. Fig.~\ref{f:J1829+4844_centroid} shows what kind 
of changes in optical centroid may happen, provided these factors are 
negligible. Whether these factors are actually negligible, we do not know.

\section{Effect of source flares}

  Rapid and strong variability on time scales from decades to weeks is 
a distinctive intrinsic characteristics of quasars. Most AGNs with 
parsec-scale jets are flaring objects. An optical variability at a level 
of 0.3 mag is rather common, and many sources exhibit changes exceeding 
one magnitude. \citet{r:smith09} provides a large numbers of light 
curves for many AGNs collected by the Steward Observatory 
spectropolarimetric monitoring project\footnote{Project website: 
\href{http://james.as.arizona.edu/~psmith/Fermi}{http://james.as.arizona.edu/\char`\~psmith/Fermi}}.
The position of the optical centroid is the weighted mean of the position 
of the starlight centroid, the accretion disk centroid, the core centroid, 
and the jet centroid, provided these components are within 
the \Gaia\ PSF. Since during a flare the brightness of only one component 
increases, the ratio of fluxes of the components changes, and the centroid 
is shifted. It matters in what direction the optical centroid is shifted 
with respect to the core. Let us denote projections of the \Gaia\ position 
with respect to the VLBI position on the jet direction $O_\mathrm{j}$ and 
on the direction transverse to the jet $O_\mathrm{t}$. 


  To what extent may $O_\mathrm{j}$ observable change due to a flare? Let us 
consider a source with the jet centroid shifted with respect to the jet base 
at 10~mas and the flux of the jet being 20\% of the total flux. According to 
expression \ref{e:e3}, the source centroid is shifted at 2~mas with respect 
to the core. If the core flux increases by 1~mag, then $O_\mathrm{j}$ becomes 
0.74, i.e., decreases by $1.26$~mas. If the core flux decreases
by 1~mag, then $O_\mathrm{j}$ becomes 3.33~mas, i.e., increases by $+1.33$~mas. 
In general, changes of optical core flux by a factor of two will cause a change 
in the positional offset of the centroid by a factor of 1.5--3. Optical 
flux changes of a factor of 2, i.e., 0.75~mag, are quite common. Analysis of the 
correlation of radio/optical polarization \citep{marscher_etal08,marscher_etal10} 
suggests that most probably, these changes happen in the compact optical core 
at parsec scales. Therefore, we conclude that $O_\mathrm{j}$ changes are observable 
and the magnitude of the change may be close to 100\% of $O_\mathrm{j}$ value of 
the quiet state.

  The sign of the change is important. There are six possible cases
(see Fig.~\ref{f:dyn}): 

{\renewcommand{\theenumi}{\arabic{enumi})}
\begin{enumerate}
    \item positive projection increases by modulo (\Ojpi);
    \item negative projection increases by modulo (\Ojmi);
    \item positive projection decreases by modulo (\Ojpd); 
    \item negative projection decreases by modulo (\Ojmd);
    \item positive projection is stationary (\Ojpz);
    \item negative projection is stationary (\Ojmz).
\end{enumerate} }

  In the first two cases we can unambiguously point in which region the flare
took place: if the positive $O_\mathrm{j}$ increases with an increase of the 
total flux density, the flare occurred in the jet. If the negative $O_\mathrm{j}$ 
projection decreases even further with an increase of the total flux density, 
the flare occurred in the accretion disk. 

  The case \Ojpd\ can be explained in two ways: a flare either in the accretion 
disk or in the core. The case \Ojmd\ can also be explained in two ways: a flare 
either in the jet or in the core. Finally, it may happen that the centroid 
is stationary (\Ojpz, \Ojmz). That means points A, B, J coincide and the 
proposed simplified scheme cannot explain the offset. 

  We see that analyzing correlation of the $O_j$ jitter and the light 
curve, we can get very valuable {\it qualitative} information: where the 
flare happened. We will show now that we are able not only to make
a qualitative inference, but investigate milliarcsecond optical structure
{\it quantitatively}. The dependence of the position centroid on changing 
brightness of the two-component model can be easily deduced from 
expression \ref{e:e3}:
\begin{eqnarray}
   O_\mathrm{j}(y) = \Frac{O_\mathrm{j}(0) + d_x \, y}{1 + y},
   \label{e:e9}
\end{eqnarray}
   where $y=\frac{\Delta F}{F}$ is the change of the flux density because of 
a flare with respect to the initial epoch $t=0$. Inverting this expression, 
we can find the shift of centroid of the component which flux density was 
constant during the flare with respect to the flaring component and
its flux density $F_f$:

\begin{eqnarray}
   \begin{array}{lcl}
      d_x(t) & = & F(0) \; \Frac{O_\mathrm{j}(t) - O_\mathrm{j}(0)}{F(t) - F(0)} + O_\mathrm{j}(t) \,, \vex \\
      F_f(t) & = & O_\mathrm{j}(0) \; \Frac{F(0)}{d_x(t)} \,. 
   \end{array}
   \label{e:e10}
\end{eqnarray}

  The light curves and time series of $O_\mathrm{j}(t)$ provide important 
redundant information. The stability of $d_x(t)$ time series will indicate that 
neither the flaring component, nor the component with constant flux density are 
moving. A statistically significant jitter of $d_x(t)$ will indicate that 
a simple stationary model does not fit the data. A straightforward interpretation 
of such a result as the time evolution of $d_x$ is problematic. 
If the jet centroid is moving, for instance, because of a motion of 
a distinctive compact feature on the jet (blob), then its flux density is 
changing. Analysis of radio jet kinematics shows this is a typical situation. 
However, jet dynamics is spawned by a process in the core. If we assume 
that the $i$-th jet component is moving along the jet, we have to assume that 
the flux density of that component, $F^j_i$ and the flux density of the core are 
changing. Analysis of kinematics of radio jets demonstrates that 
the following simplified model works most of the time \citep{r:lister16}. 
The core ejects components at discrete epochs. After ejection, the component 
moves mainly linearly. Its flux density is zero before the ejection eoch and 
becomes zero after some time. For such a simplified model, equations 
for $O_\mathrm{j}(t)$ and the total flux density $F_\mathrm{t}(t)$ are written as

\begin{eqnarray}
   \begin{array}{lcl}
      O_\mathrm{j}(t) & = & \dss\sum_i \Frac{v(t-t_{0i}) \, F^j_i(t) \: + \:
                             d_i(t_{0i}) \, F^j_i(t_{0i})}
                              {F_c(t) \: + \: \dss\sum_k F^j_k(t)} \,,\\
      F_\mathrm{t}(t) & = & F_c(t) + \dss \sum_i F^j_i(t) \,,            \\
      F^j_\mathrm{i}(t) & = & 0 \,, \quad \forall \: t < t_{0i} \,,
   \end{array}
   \label{e:e11}
\end{eqnarray}
  where $F_\mathrm{c}(t)$ is the combined flux density of the core and 
starlight. $O_\mathrm{j}(t)$ and $F_t(t)$ are measurements, and 
$v$, $F_c(t)$, $F^j_i(t)$, $d_i(t_{0i})$, and $t_{0i}$ 
are unknowns. In general, the system does not have a unique solution, 
however using additional information may make this system solvable.

  Let us consider a system that consists of 1)~a core with variable flux
density $F_\mathrm{c}(t)$ that includes also the contribution of starlight 
and 2)~a jet component that moves with a constant angular velocity $v$ with 
variable flux density $F_\mathrm{j}(t)$ computed by integrating its 
intensity distribution. The system is observed from the moment $t_b$ that 
is not necessarily equal to the epoch of the jet component ejection $t_0$. 
For such a model the flux density of the moving jet component is 
expressed as
\begin{eqnarray}
   \begin{array}{lcl}
      F_\mathrm{j}(t) & = & \Frac{O_j(t) \, F_\mathrm{t}(t) - O_\mathrm{j}(t_b) \, F_\mathrm{t}(t_b)}
                            {v \, (t - t_b)} + F_\mathrm{j}(t_b)            \vex \\
      F_\mathrm{c}(t) & = & F_\mathrm{t}(t) - F_\mathrm{j}(t)                                 \vex \\
      d_j(t) & = & d(t_b) + v(t-t_b).
   \end{array}
   \label{e:e12}
\end{eqnarray}

   If we know the angular velocity of a component, we can determine its 
light curve, the light curve of the core, and the evolution of the component 
centroid displacement. The velocity can be derived from radio observations. 
This is an intrinsic property of a source that does not depend on frequency. 
However, expression \ref{e:e12} is applicable only for an interval of 
time when there is only one component. Determining the interval of validity 
of expression \ref{e:e12} requires utilizing additional information.

  A complication arises from the fact that the \Gaia\ position estimates
of weak objects like AGNs are almost entirely derived using the data sampled 
along the scanning direction. A \Gaia\ position at a given epoch is one-dimensional.
Therefore, at a given time epoch the uncertainties of $O_\mathrm{j}$ 
and $O_\mathrm{t}$ depend on the angle between the scanning direction and the jet 
direction. At some epochs $O_\mathrm{j}$ or $O_\mathrm{t}$ observables may 
have so large uncertainties what will make them unusable for parameter 
estimation. Since the scanning direction changes with time due to the \Gaia\ orbit 
precession, the uncertainties of the mean $O_\mathrm{j}$ and $O_\mathrm{t}$ 
observables mainly do not depend on scanning direction.

  We should notice that the effect of source variability on position 
changes of objects with structure confined within the PSF is not new. It was 
discussed before, \citep[f.e.,][]{r:wielen96,r:jayson16} in relation to the 
HIPPARCOS and USNO-B1.0 catalogues. As it was shown by \citet{r:wielen96}, 
time series of only the total flux and position displacements are 
sufficient for establishing the system has a structure, f.e.\ whether the object
is binary, but are not sufficient for a separation of variables and determination 
of the distance between the components and their flux densities. In contrast, 
using $O_\mathrm{j}$ observables permits variable separation in a case of 
a simple structure, since it is based on additional information: VLBI position 
of the core.

\section{Jitter in Gaia source position estimates and mitigation of 
its impact}

  An inevitable consequence of interpretation of the observed VLBI/\Gaia\ 
position differences as a manifestation of the optical jet is 
the non-stationarity of the centroid position determined by \Gaia.
Brightening of the core and, possibly, the accretion disk causes 
non-stationarity of the centroid. Jet kinematics, i.e., appearance and
motion of new features in the jet, their motion and intensity evolution 
influences the position of the centroid as well. Both processes are stochastic 
and non-predictable. Therefore, we call 
it rather a jitter than a proper motion. A change in apparent position
of \Gaia\ centroids due to these processes differs from a motion of stars 
that is a combination of the motion in the Galactic gravity field, the 
orbital motion for binary or multiple system, and gravitational 
bending. \citet{r:lar17} showed that micro-lensing due to randomly moving
point masses in the gravitational field of the Galaxy will cause random 
noise in apparent position of objects located within the Galactic plane 
at a level of tens microarcseconds, but above that the level proper motion 
of stars is regular. Although proper motion of SMBH is expected to be 
negligible at least at the level of microarcseconds, the position of the 
\Gaia\ centroid may change at the level of milliarcseconds. This change 
is irregular and unpredictable.

  The instability of AGN position estimates derived from VLBI observations 
was known for a long time \citep[f.e.,][]{r:gont01}. This instability 
is related to the omitted term $\tau_\mathrm{s}$ that accounts for source
structure in data reduction. Scattering of radio emission in the 
interstellar medium also changes apparent radio images and may increase 
the errors of VLBI position estimates. This effect is most prominent in 
the Galactic plane \citep[e.g.,][]{r:pushkarev13,r:puskov15}.

  The discovery of the presence of optical jets from VLBI/\Gaia\ comparison 
by \citet{r:gaia2} raises the problem of the source position jitter in the 
optical range. However, optical jets contribute to the centroid position 
differently. First, as we see from Table~\ref{t:bl229aa}, position of the 
image centroid is more sensitive to the extended jet structure than 
the position derived from group delays. Second, the centroid position is 
sensitive not only to the motion of a jet component or its brightening, 
but more importantly to well known strong variability of the optical emission 
of the core or even the accretion disk without changes in the jet.

  Absolute astrometry catalogues based on star observations are marred
by errors that originate from uncertainties of star proper motions,
which sets the limit of a catalogue accuracy \citep[e.g.,][]{r:sovers}. 
The position accuracy degrades with time since the contribution of uncertainties 
in proper motions to source positions at a current epoch accumulates with 
time. Remote galaxies that are located so far what makes their transverse 
motion negligible were considered for a long time as ideal targets that are 
supposed to eliminate this problem \citep{r:wright50}. The reality turned 
out different. Analysis of VLBI results showed that the problem of degrading
position accuracy with time has gone, but a new problem appeared: position 
jitter due to extended parsec-scale variable structure that affects position 
estimates. We predict a similar situation in the optical range, even at 
a larger scale.

  The problem of the source position jitter in VLBI results can be alleviated 
by changing scheduling and analysis strategy. If observations are scheduled
and calibrated in such a way that they can be used for generating source 
images, then $\tau_\mathrm{s}$ term can be computed and applied in data 
analysis. \citet{r:cha02} has demonstrated reduction of the source position 
scatter using this approach to a limited data set. Applying source structure 
for processing the observations collected under absolute astrometry and geodesy 
VLBI programs has not yet become common because it requires significant efforts
and promises a little return: improvement in the source position stability at 
a level of a tenth of a milliarcsecond has a negligible effect on estimates of 
Earth orientation parameters or station positions \citep{r:xu16} with respect to 
other systematic errors and it is small with respect to typical thermal noise 
in source positions \citep[0.5~mas among VLBI/\Gaia\ counterparts,][]{r:gaia1}.

  In a similar way, the problem of a source jitter in the optical centroid 
positions can be alleviated. First, we expect position variations 
to be not totally random. The position jitter will have a preferable 
direction along the jet, as it was established from analysis of VLBI/\Gaia\ 
position offsets \citep{r:gaia2}. Analysis of radio jet kinematics shows 
that transverse jet motions are rare \citep{r:lister16}. While we expect some 
jitter in source positions along the jet, we expect the jitter in the 
transverse direction to be significantly less and probably not detectable 
with \Gaia. Second, we expect the correlation between the centroid position 
jitter and the flux changes in the optical range. The larger the flux
density variations, the larger the expected centroid position jitter.

  Jet directions can be determined from radio observations of radio-loud
AGNs. For AGNs which lack information on their jet direction from VLBI 
images the jet direction can be determined from analysis of their \Gaia\ 
centroid time series. The scatter of the source positions in a plane 
tangential to the source direction can be described  by a sum of two 
distributions: the 2D Gaussian distribution associated with errors in
position time series and the distribution of the source position wander 
along a certain direction due to the presence of the optical jet.
Fitting a straight line into the two-dimensional scatter of source position
estimates with respect to the weighted mean will allow us to restore the jet
direction. Since the error ellipse of \Gaia\ positions at each individual 
epochs is strongly elongated across the scanning direction, the distribution
of scanning directions determines whether the jet direction can be determined.
If the distribution of scanning directions is substantially non-uniform, 
a reliable determination of jet direction even in the presence of jitter 
is problematic.

  Analysis of $O_\mathrm{j}$ observables time series and optical fluxes 
may in some favourable cases allow us to determine the position of the optical 
core. If the optical jet of a two-component core-jet model is stable, which can 
be deduced from stability of $d_x(y)$ time series in expression \ref{e:e10}, 
then using the mean value of $d_x(y)$ and jet direction from VLBI, we will get
a precise position of the optical core, which is different than the mean 
position of the centroid. If $d_x(y)$ time series show no systematic changes, 
determination of the optical core is possible. Since the denominator in 
expression \ref{e:e10} has the variation of the optical flux with respect to 
the flux at the initial epoch, the accuracy of the optical core determinations 
is higher when the optical flux variations are higher. Thus, the synergism of 
VLBI and \Gaia\ allows us in these cases to alleviate the contribution of the 
jitter of the centroid position, solve for the VLBI/\Gaia\ bias, and determine 
position of the optical core. If the number of sources for which the position of 
the core can be determined will be high enough, these sources can be used for
improvement in determination of the orientation and drift of the \Gaia\ 
catalogue.

  Assuming AGN position estimates are stable in time, the orientation and 
drift of the \Gaia\ catalogue can be characterized by three parameters. 
Rotation angles, can be computed assuming the net rotation in VLBI and 
\Gaia\ positions among matching sources is zero 
\citep[See eq.~5 in][]{r:gaia_dr1}.

  A small rotation that can be represented as vector $\vec\Psi$ with
Cartesian coordinates $\Psi_1, \Psi_2, \Psi_3$ applied to an objects
with polar coordinates $\alpha, \delta$ will cause increments in coordinates
$\Delta \alpha, \Delta \delta$:

\begin{eqnarray}
   \begin{array}{rcrrr}
       \Delta \alpha & = & -\cos \alpha \tan \delta \: \Psi_1 & -\sin \alpha \tan \delta \: \Psi_2 & + \: \Psi_3 \\
       \Delta \delta & = &              \sin \alpha \: \Psi_1 &             -\cos \alpha \: \Psi_2 &          \\
   \end{array}
   \label{e:e13}
\end{eqnarray}

  The coordinates of the rotation vector can be determined with least squares 
requiring that the position difference of matching sources with respect
to VLBI be zero. In absence of the jitter, the reciprocal weights of observation 
equations are 
$1/w_\alpha = \sqrt{\sigma_\mathrm{v}^2 + \sigma_\mathrm{g}^2}\cos\delta$ for 
right ascensions and $1/w_\delta = \sqrt{\sigma_\mathrm{v}^2 + \sigma_\mathrm{g}^2}$ 
for declinations, where $\sigma_\mathrm{v}$ and $\sigma_\mathrm{g}$ are 
uncertainties in VLBI and \Gaia\ positions. In order to take into account 
the jitter, we just inflate the position uncertainties along the jet direction:

\begin{eqnarray}
   \begin{array}{lcl}
      1/w_\alpha $ = $ \sqrt{\sigma_{\alpha,v}^2 + \sigma_{\alpha,g}^2 + \sigma_j^2 \sin^2 p} \, \cos\delta \vex \\
      1/w_\delta $ = $ \sqrt{\sigma_{\delta,v}^2 + \sigma_{\delta,g}^2 + \sigma_j^2 \cos^2 p},
   \end{array}
   \label{e:e14}
\end{eqnarray}
   where $\sigma_\mathrm{j}$ is the second moment of the jitter distribution along
the jet and $p$ is the jet positional angle. Precise knowledge of 
$\sigma_\mathrm{j}$ is not important. Selecting 
$\sigma_\mathrm{j} \gg \max(\sigma_\mathrm{v},\sigma_\mathrm{g})$ will effectively 
down-weight the projection of the position difference along the jet, and 
the estimation process will use only the transverse projection in solving 
system \ref{e:e13}.

\section{Galaxies with weak jets}

  We should refrain from a generalization of results of our analysis of 
VLBI/\Gaia\ offsets of the AGNs detected with VLBI to the entire population of 
active galaxies.  The population of the AGNs selected on this basis of their 
parsec-scale radio emission with the cutoff at 10 mJy at 8~GHz is biased towards 
relativistically-boosted jets with small viewing angles 
\citep[e.g.,][]{Cohen_etal07,r:Hovatta_etal09,r:pushkarev17} 
resulting in the effects reported by \cite{r:gaia2} and discussed in this paper. 
\citet{r:kel16} showed that for roughly 80\% objects in the complete 
optically-selected sample of quasars their 6~GHz radio emission from
star-forming regions dominates, rather than from the synchrotron radiation 
of jets. Since emission from star-forming regions is much weaker, these
objects are radio-quiet. Thus, the majority of the \Gaia\ AGNs that are
selected on the basis of their optical flux with the cutoff at $20.7^m$ are 
radio-quiet with radio emission from jets extremely weak or even absent. 
Considering argumentation of \citet{r:perlman2010} that radio and optical 
jet emission is caused by the same mechanism, we conclude that optical jets 
of the radio-quiet AGNs sample are expected to be also extremely weak or even 
absent. At the same time, previous studies have demonstrated 
\citep[see, e.g.,][]{Elvis_etal94,KB99,SOS04} that optical emission of 
the accretion disk and/or the host galaxy dominates for the population of 
AGNs selected on the basis of their optical fluxes. Consequently, 
the \Gaia-selected AGNs should have a much smaller share of objects with 
significant emission of the jet than the VLBI-selected ones.

  If to exclude emission from the optical jet and consider only 
the contribution from the accretion disk and from the starlight of the host 
galaxy, the optical centroid position will be affected by the displacement of 
the starlight centroid with respect to the accretion disk. For galaxies 
that do not interact with nearby companions and have no asymmetries, such as
dust bars, these two points are expected to be very close, and the accretion disk 
variability should cause very small centroid displacements. Though, \citet{r:pop12}
argue that perturbations in the inner structure of the accretion disk and 
surrounding dusty torus may reach a milliarcsecond level for luminous AGNs
at small redshifts. To which extent these points are close, will be seen from 
analysis of the correlation of light curves with position time series.

  In general, the positions of the radio-quiet AGNs are expected to be
more stable than the positions of the radio loud sample since the contribution
of one of the factors that affects position stability, the optical jet, is 
excluded. The position accuracy of the radio-quiet AGN sample may be the 
higher the position accuracy of the radio-loud AGN sample, but unfortunately, 
currently there is no practical way to obtain precise coordinates of such 
objects with VLBI and use them for radio/optical ties. In this context, the 
distinction between two AGNs populations is drawn based on whether the 
synchrotron emission dominates in the total flux density (radio-loud) or 
not (radio-quiet).

\section{Future observations}

   Before the \Gaia\ launch, it was considered for a long time that the main
obstacle for VLBI/\Gaia\ comparison would be a small number of 
suitable extragalactic radio sources. Dedicated programs for VLBI observations 
of several hundreds new suitable candidates for matching the catalogs \citep{r:bourda11} 
or improving positions of several hundreds known sources \citep{r:bail16} were 
made. It was expected that these efforts will significantly help to align the VLBI 
and the \Gaia\ source position catalogues and investigate zonal errors of the 
catalogues.

   The \Gaia\ data release followed by the discovery of significant 
contribution of extended optical structure in \Gaia\ positions \citep{r:gaia2} 
had a profound impact. First, it was found that roughly one half of the VLBI 
sources have a \Gaia\ counterpart that has a weak dependence on radio
flux density \citep[Fig.~1 in][]{r:gaia1}. A dedicated search of new \Gaia\ 
counterparts does not seem to be necessary. Any VLBI survey will increase 
the number of VLBI/\Gaia\ matches with a rate of about one match per 
two-three new sources. By August 1, 2017 the total number of compact radio 
sources detected with VLBI under absolute astronomy programs reached 14,767.
Among them, there are 7669 matches with \Gaia\ with the probability of false 
association less them $2 \cdot 10^{-4}$. There will be no problem related to 
a shortage of matching sources for VLBI/\Gaia\ comparison, and the comparison 
itself will not be limited to an alignment of catalogues and studying zonal 
errors.


  As we have shown, VLBI/\Gaia\ position differences bring invaluable information. 
The value of this information is significantly enhanced if the jet direction is 
known and we can derive $O_\mathrm{j}$ and $O_\mathrm{t}$ observables. \Gaia\ will 
provide time series of source positions accompanied by light curves. Analysis 
of $O_\mathrm{j}(t)$, $O_\mathrm{t}(t)$ time series and light curves will be 
a powerful tool probing optical jets at scales two order of magnitude finer 
than the resolution of current and perspective optical telescopes. Under best
conditions with no more than one evolving component, combined analysis
of VLBI and \Gaia\ will be able to provide the evolution of optical jet 
centroids at milliarcsecond scales.

  In order to make such a deep insight into optical structure, VLBI has to
solve several problems. VLBI positions of all the matches should be determined
with accuracy not worse than the accuracy of \Gaia. High quality radio 
images of matching sources should be produced. This will allow us to compute
the source structure contribution and apply a correction during data reduction.
Directions of jets have to be determined. We do not know in advance when 
a given source will have a flare. Therefore, it is desirable to have this 
information for all the matches (about~8,000). At the moment, the 
median accuracy of the VLBI position catalogue {\sc rfc\_2017a}\footnote{Available
at \href{http://astrogeo.org/rfc}{http://astrogeo.org/rfc}} 
(Petrov \& Kovalev, in prep.) is 0.8~mas, while 22\% of the sources have 
position errors exceeding 2~mas because of the thermal 
noise. Technically, using observations at VLBA or other large VLBI arrays, we 
can determine source positions with accuracy better than 0.2~mas if a given 
source is observed long enough. According to our analysis, systematic errors 
dominate beyond the 0.2~mas accuracy level. 

  In the past, there was no strong demand to have high position accuracy 
for all the sources with term $\tau_\mathrm{s}$ applied in data analysis and 
have their high fidelity images. At the moment, source images are available 
for 80\% objects observed under absolute astrometry 
programs\footnote{See \href{http://astrogeo.org/vlbi_images}{http://astrogeo.org/vlbi\_images}}.
Of them, jet directions can be reliably determined for one half of the objects 
with an automatic procedure \citep{r:gaia2}. Source images for 4412 objects
(47\%) were derived from 60~s long snapshot observations made in one scan, 
which is not sufficient for achieving high imaging quality. Observing sources
longer, in 3--6 scans, will increase the share of images where we can 
determine jet direction to over 90\%. We should stress that all these listed 
problems can be solved with existing facilities under dedicated program. At 
the same time, attempts to add some sources to regular geodetic VLBI 
observations \citep{r:bail16,r:shu17} turned out only partly successful. 
Improvement of source position coordinates with a pace of 30--100 sources per 
year is not sufficient to make a noticeable difference. Therefore, we envisage
dedicated programs targeting all 8000 matches. The focus of these programs 
will be shifted from densification of the VLBI catalogue and finding suitable 
matches to refining source positions and images. 

  Such a large dataset of precise determinations of $O_\mathrm{j}$ and 
$O_\mathrm{t}$ observables will be useful for a number of applications. 
First, the time series of $O_\mathrm{j}(t)$, $O_\mathrm{t}(t)$ accompanied 
with light curves and, if available, with a series of radio images, will 
be useful for deriving a model of optical jet evolution of objects of 
interest. $O_\mathrm{t}(t)$ observable will be useful for evaluation of 
random and systematic errors not related to the presence of  optical 
structure. When the noise in the differences due to other factors
affecting VLBI/\Gaia\ positions is small with respect to $O_\mathrm{j}$, 
individual sources can be studied. 
  
   Second, the bulk data of mean values and standard deviations of these 
observables will be used for statistical studies correlating $O_\mathrm{j}$
and its evolution with other properties of AGNs. Statistical studies are 
possible even when accuracy of $O_\mathrm{j}$ observables is low and 
not sufficient for analysis of individual sources.

  Third, a population of AGNs without radio counterparts can be 
studied. The jet direction can be found from the analysis of a scatter of 
position time series. The sources with significant asymmetry in their 
two-dimensional position scatter should be considered as candidates to 
AGNs. Correlation between $O_\mathrm{j}$ and the position jitter makes 
classification of a given source as an AGN almost certain.

  Statistical analysis of $O_\mathrm{j}(t)$ and light curves has a potential to answer
a number of interesting questions, such as how often, if ever, do flares occur 
in the accretion disk area; how often do flares occur in jet components; how 
long typical optical jets are; what is the role of jet kinematics in a jitter
of optical centroids and what is the role of core variability.

\section{Summary and conclusions}

   Analysis of VLBI/\Gaia\ positional offsets revealed they are not entirely 
random \citep{r:gaia1}. The presence of a preferable direction in the 
distribution of the offsets firmly associates them with an intrinsic property 
of AGNs: core-jet morphology \citep{r:gaia2}. Since VLBI records voltage that 
is later cross-correlated and \Gaia\ uses a quadratic detector, the CCD camera, 
the response of the instruments to source structure is fundamentally 
different. We have simulated, tested, and confirmed that VLBI is sensitive 
mainly to the position of the most compact detail, the AGN core. With a proper 
analysis procedure, the effect of source structure on position estimate can 
be reduced to below the 0.1~mas level. The contribution of the optical source 
structure on the centroid position derived from \Gaia\ is usually greater 
due to a higher weight of the extended low surface brightness emission.

   We predict a jitter in \Gaia\ centroid position estimates
for radio loud AGNs. It is mainly caused by variability of the optical 
core flux density relative to the slowly varying jet. The magnitude of the 
jitter depends on the magnitude of flux density variations and the extension 
of the jet. For highly variable sources it may reach several milliarcseconds. 
The presence of an unpredictable jitter in source positions is already known 
in VLBI astrometry results, but is new in the field of optical space astrometry. 
The radio-quiet AGNs may be more suitable for construction of a highly 
precise optical reference frame since they are expected to have more stable 
optical positions.

  Using accurate astrometric VLBI position as a reference point of the stable 
radio jet base in an AGN, we can form new observables $O_\mathrm{j}$ and $O_\mathrm{t}$~--- 
projections of the VLBI/\Gaia\ position difference on the parsec-scale 
jet direction and 
the direction transverse to the jet. We have shown that these observables 
and the optical light curves are a powerful tool for studying optical jets 
at the milliarcsecond scales, unreachable for any other instrument. Analysis of 
$O_\mathrm{j}(t)$ time series and optical light curves may allow recovering 
properties of the optical core-jet morphology: position of the jet centroid, 
its flux density, and in some simple cases kinematics. Analysis of these series 
has a potential to locate the region where the optical flare occurs: in the 
core, the accretion disc, or jet features.

  A recognition of the fact that optical positions of radio loud AGNs 
cannot be considered as point-like unmovable sources at the \Gaia\ level 
of positional accuracy leads to a paradigm shift in the field of high precision 
absolute astrometry.\
%
%
%

  The presence of optical structure at 1--2~mas level associated with 
relativistic jets revealed in the early \Gaia\ data release for VLBI-selected 
AGNs sets the limit to which extent \Gaia\ positions can  be used for radio 
astronomical applications. At the accuracy level worse than that threshold,
\Gaia\ positions can be used for radio astronomy and vice versus. At the 
accuracy level better than that threshold, the positions divert since VLBI 
and \Gaia\ ``see'' different parts of a complex radio-loud AGN with a bright 
relativistically-boosted jet. That means a single technique cannot produce 
the reference frame that is suitable for every wavelength range even in 
principle. The \Gaia\ DR1 has already surpassed that accuracy threshold. 
Further improvement in position accuracy of VLBI and \Gaia\ will not results 
in a reconciliation of radio and optical positions but will results
in improvement of accuracy of determination of these position differences. 
The differences are not solely due to errors in position estimates, but 
contain a valuable signal. Investigation of this signal will belong to the 
realm of astrophysics.

  The applications that require positions of radio objects with 
accuracy better than 1--2~mas, such as space navigation, Earth orientation 
parameter measurement, determination of the orientation of the Earth's orbit 
from combined analysis of pulsar positions from VLBI and timing, cannot 
borrow coordinates of observed objects from \Gaia, but will have to rely on
their determination from VLBI in the foreseeable future.

\section*{Acknowledgments}

It is our pleasure to thank Claus Fabricius and Eduardo Ros for 
a thorough review of the manuscript and valuable suggestions that
have helped to improve the manuscript. We would like to thank Sergei
Sazonov and Ian Browne for fruitful discussions.

This project is supported by the Russian Science Foundation grant 16-12-10481.
This work has made use of data from the European Space Agency (ESA) mission 
Gaia\footnote{\href{https://www.cosmos.esa.int/gaia}{https://www.cosmos.esa.int/gaia}}, 
processed by the Gaia Data Processing and Analysis Consortium 
(DPAC\footnote{\href{https://www.cosmos.esa.int/web/gaia/dpac/consortium}{https://www.cosmos.esa.int/web/gaia/dpac/consortium}}). 
Funding for the DPAC has been provided by national institutions, in particular 
the institutions participating in the Gaia Multilateral Agreement.
This research has made use of data from the MOJAVE database that is maintained by
the MOJAVE team \citep{r:lister09}
Some of the data presented in this paper were obtained from the Mikulski 
Archive for Space Telescopes (MAST). STScI is operated by the Association of 
Universities for Research in Astronomy, Inc., under NASA contract NAS5-26555. 
Support for MAST for non-HST data is provided by the NASA Office of Space Science 
via grant NNX09AF08G and by other grants and contracts.
%
%
We used in our work VLBA data provided by the Long Baseline Observatory that 
is a facility of the National Science Foundation operated under cooperative 
agreement by Associated Universities, Inc.

\bibliographystyle{mnras}
\bibliography{gaia3}

\bsp    
\label{lastpage}
\end{document}